\documentclass[11pt,a4paper]{article}
\usepackage[margin=1in]{geometry}
\usepackage[utf8]{inputenc}
\usepackage{xcolor}
\usepackage{fancyhdr}
\usepackage{amsmath,amssymb}
\usepackage{multirow,booktabs,graphicx}
\usepackage{mathtools,feynmp-auto}
\usepackage{multirow}
\usepackage{tikz}
\usepackage{pgfplots}
\usepackage{placeins}
\usepackage{cite}
\usepackage[normalem]{ulem}

\usepackage{xspace}
\usepackage[shortlabels]{enumitem}
\usepackage{jheppub}

\graphicspath{{figures/}}

\def\tablename{Table}
\def\figurename{Figure}

\newcommand\one{\leavevmode\hbox{\small1\normalsize\kern-.33em1}}

\newcommand{\mev}{\text{MeV}}
\newcommand{\gev}{\text{GeV}}

\newcommand{\TP}{\text{TP}}
\newcommand{\FP}{\text{FP}}

\newcommand{\TPR}{\text{TPR}}

\newcommand{\NNone}{NN$_\text{tight}$\xspace}
\newcommand{\NNtwo}{NN$_\text{loose}$\xspace}
\newcommand{\NNbinned}{NN$_\text{binned}$\xspace}

\newcommand{\AUC}{\text{AUC}}

\def\slashchar#1{\setbox0=\hbox{$#1$}           
   \dimen0=\wd0                                 
   \setbox1=\hbox{/} \dimen1=\wd1               
   \ifdim\dimen0>\dimen1                        
      \rlap{\hbox to \dimen0{\hfil/\hfil}}      
      #1                                        
   \else                                        
      \rlap{\hbox to \dimen1{\hfil$#1$\hfil}}   
      /                                         
   \fi}

\renewcommand{\Vec}{%
  \mathpalette {\overarrow@\vectfill@}}
\def\vectfill@{\arrowfill@\relbar\relbar{\raisebox{-3.81pt}[\p@][\p@]{$\mathord\mathchar"017E$}}}

\newcommand{\be}{\begin{eqnarray*}}
\newcommand{\ee}{\end{eqnarray*}}

\newcommand{\bee}{\begin{eqnarray}}
\newcommand{\eee}{\end{eqnarray}}
\newcommand{\beeq}{\begin{equation}}
\newcommand{\eeeq}{\end{equation}}

\newcommand{\vub}{\ensuremath{|V_{ub}|}\xspace}

\newcommand{\Sherpa}{S\protect\scalebox{0.8}{HERPA}\xspace}
\newcommand{\Pythia}{P\protect\scalebox{0.8}{YTHIA8}\xspace}
\newcommand{\Photos}{P\protect\scalebox{0.8}{HOTOS}\xspace}
\newcommand{\Evtgen}{E\protect\scalebox{0.8}{VT}G\protect\scalebox{0.8}{EN}\xspace}

\begin{document}
\thispagestyle{empty}
\begin{flushright}
IPPP/20/104 \\
\end{flushright}
\vspace{0.8cm}

\begin{center}

\begin{center}


{\Large\sc Potential and limitations of machine-learning approaches to inclusive $|V_{ub}|$ determinations}

\end{center}

\vspace{0.8cm}

\textbf{
Anke Biek\"otter$^{\,a,b}$, Ka Wang Kwok$^{\,a,b}$ and Benjamin D.~Pecjak$^{\,a}$}\\

\vspace{1.cm}
{\em {$^a$Institute for Particle Physics Phenomenology, Department of Physics, Durham University, South Road, 
Durham DH1 3LE, United Kingdom}}\\

{\em {$^b$Institute for Data Science, Durham University, South Road, 
Durham DH1 3LE, United Kingdom}}\\[0.2cm]

{}
\vspace{0.5cm}


\end{center}

\begin{abstract}

The determination of $|V_{ub}|$ in inclusive semileptonic $B \to X_u \ell \nu$~decays will be among
the pivotal tasks of Belle~II. 
In this paper we study the potential and limitations of machine-learning approaches that attempt
to reduce theory uncertainties by extending the experimentally 
accessible fiducial region of the $B\to X_u \ell\nu$ signal into regions where the
$B\to X_c \ell\nu$ background is dominant. We find that a deep neural network trained on low-level single particle features offers modest improvement in separating signal
from background, compared to BDT set-ups using  
physicist-engineered high-level features.
We further illustrate that while the signal
acceptance of such a deep neural network deteriorates in kinematic
regions where the signal is small, such as at high hadronic invariant mass, 
neural networks which exclude kinematic features are flatter in kinematics but
less inclusive in the sampling of exclusive hadronic final states at fixed kinematics.
The trade-off between these two set-ups is somewhat Monte Carlo dependent, 
and we study this issue using the multipurpose event generator \Sherpa
in addition to the widely used $B$-physics tool \Evtgen.  
\end{abstract}

\newpage

\tableofcontents

\newpage
\section{Introduction}
\label{sec:intro}

The determination of the parameters of the CKM matrix is an 
important test of the Standard Model (SM). Its least known element is $\vub$, which
can be determined at $B$-factories from semileptonic $B$-decays in the exclusive $B \to \pi \ell \nu$~channel~\cite{Ha:2010rf,delAmoSanchez:2010af,Lees:2012vv,Sibidanov:2013rkk}  
as well as from inclusive  
$B \to X_u \ell \nu$~decays~\cite{Urquijo:2009tp,Lees:2011fv,Cao:2021xqf}. 
Moreover, it can be tested at the LHCb experiment in
$\Lambda_b \to p \mu \nu_\mu$ decays~\cite{Aaij:2015bfa}.
The current average value of inclusive and exclusive measurements is 
$\vub = (3.82 \pm 0.24)\times 10^{-3}$~\cite{Tanabashi:2018oca}.
However, there is a long-standing $~3\sigma$~tension between them, 
making  the determination of $\vub$ in the inclusive mode an exciting future measurement for Belle~II.

From the theoretical standpoint, the total $B \to X_u \ell \nu$ decay rate would offer 
the cleanest extraction of $\vub$. It can be calculated
using the local operator product expansion (OPE) familiar from inclusive semileptonic 
decay into charm quarks, $B \to X_c \ell \nu$ \cite{Chay:1990da, Bigi:1992su, Blok:1993va, Manohar:1993qn}. At leading order in this $1/m_b$ expansion the result for the inclusive
decay is equal to that for the quark-level process $b\to u \ell \nu$,
whose total \cite{vanRitbergen:1999gs} and differential  \cite{Brucherseifer:2013cu} decay rates are known up next-to-next-to-leading order in QCD.  At 
relative order $1/m_b^2$ only a handful of non-perturbative parameters appear, and 
recently even for these power corrections the next-to-leading-order
QCD corrections have been calculated~\cite{Capdevila:2021vkf}. 

From the experimental standpoint, the large background from charmed final states precludes a straightforward measurement of the total inclusive $B \to X_u \ell \nu$ decay rate.  The traditional approach to inclusive $\vub$ measurements has thus been to make kinematic cuts to restrict measurements in phase-space regions which, neglecting detector effects, are free from charm background. Examples of such cuts are 
$M_X < m_D$, where $M_X$ is the invariant mass of the hadronic final state $X$ and $m_D$
is the $D$-meson mass, or $P_+ < m_D^2/m_B$, where
$P_+ = E_X- |\vec{P}_X|$ is the energy-momentum  difference of the hadronic 
final state  and $m_B$ is the $B$-meson mass.  
 
Even apart from the fact that detector effects cause the charm background to populate these theoretically charm-free phase-space regions (see  Figure~\ref{fig:mass_detector_yes_no} below), requiring a non-trivial separation of  signal and background also for these restrictive kinematic cuts,  the theoretical description of the partial  $B \to X_u \ell \nu$  decay rates becomes  considerably more involved. To the extent that the phase-space cuts limit the partial decay rates to 
the ``shape function region'',  where the hadronic final state is a collimated jet whose energy is much larger than its invariant mass, the local OPE breaks down and is replaced by a non-local, shape function OPE.  
The leading-order contribution in the corresponding $1/m_b$ expansion involves a single non-perturbative shape function \cite{Neubert:1993ch, Bigi:1993ex}, which is a function of one light-cone variable.  Analyses in soft-collinear effective theory have shown that  the $1/m_b$ power corrections in this non-local OPE involve a plethora of subleading shape functions beyond tree level, some of which are a function of up to three light-cone variables~\cite{Lee:2004ja, Bosch:2004cb, Beneke:2004in}, and that the next-to-next-to-leading order QCD corrections to the leading-power decay rate can be substantial~\cite{Greub:2009sv}.  

Phenomenologically, several theoretical approaches to partial $B \to X_u \ell \nu$ decay rates are used in $\vub$ extractions, going under the acronyms ADFR~\cite{Aglietti:2008hd}, BLNP~\cite{Bosch:2004th,Lange:2005yw}, DGE~\cite{Andersen:2005mj} and GGOU~\cite{Gambino:2007rp}.  These differ in the treatment of QCD effects
in the shape function region, but all reduce to the conventional, local OPE results if the kinematic cuts do not introduce new scales which are parametrically much smaller than the $b$-quark mass.  
Given the complicated structure of the factorisation theorems, the debate over the precise nature of the  shape-function OPE, and the fact that there is no obvious new physics explanation for the current discrepancies between inclusive and exclusive determinations \cite{Crivellin:2014zpa}, it is clearly desirable to extend measurements over as large a region of phase space as possible, such that the theoretically clean local OPE results can be applied.

Multivariate analysis techniques based on machine learning (ML)   
are ideally suited for accessing the $B\to X_u\ell \nu$ decays in 
regions dominated by the $B\to X_c\ell \nu$ background,  while still achieving 
good signal-to-background ratios. From the ML perspective, the 
challenge is to build a classifier between signal ($B\to X_u\ell \nu$) and
background ($B\to X_c\ell \nu$ and other decays).  
The first example of such a ML approach to $\vub$  determinations was the Belle analysis of Ref.~\cite{Urquijo:2009tp}. It used a boosted decision tree~(BDT) based classifier taking various high-level kinematic and global features as input and gave a result for the partial decay rate with the single restriction that the charged lepton carries momentum greater than $1\,\gev$ in the $B$-meson rest frame. Thereby, it samples more than 90\,\% of the inclusive $B\to X_u\ell \nu$ phase space such that a theoretical description based on the local OPE is applicable. A potential criticism is that such a classifier needs to be trained on Monte Carlo (MC) samples of signal and background events, and is thus especially susceptible to systematic errors based on the kinematic modelling of the signal.  A possible approach 
to evading this criticism was presented recently in the reanalysis of the Belle data in Ref.~\cite{Cao:2021xqf}, where kinematic properties  were not included as input features in a BDT classifier. 
Although the classification power of such a BDT is reduced when viewed in terms of typical machine-learning metrics such as the area under the curve,
it can be used to enhance the signal-to-background ratio to a level which permits a binned one- and two-dimensional likelihood analyses of the kinematic features of the signal and background after event selection resulting in a similar significance after the likelihood analysis.

The purpose of this paper is to perform a systematic study on the use of ML-based classifiers for inclusive $\vub$ analyses.  We focus on two 
main aspects. First, we explore
the use of deep neural nets (NNs) as an alternative ML architecture to BDTs.  While BDTs
typically work best when given a small set of carefully engineered, high-level features 
such as the hadronic invariant mass, NNs can take as input the very 
high-dimensional set of low-level features characterising the event (such as the four-momenta of the final-state particles) and use it to learn an optimal way to classify 
signal and background.\footnote{For some discussions on the benefits of using 
low-level features rather than expert engineered high-level input features only, see e.g.~Refs.~\cite{Baldi:2014kfa, Guest:2016iqz} or the ML review~\cite{Guest:2018yhq}. }
Second, we study in detail the inclusivity of the classifiers and their sensitivity not only to the set of input features chosen, but also to the event generator used producing the training data.
In particular, while present $\vub$ analyses rely on 
the generator \Evtgen~\cite{Lange:2001uf},  in this paper we compare results using combinations of 
\Sherpa~\cite{Gleisberg:2008ta} and \Evtgen event samples, which differ very little in their description of the $B\to X_c \ell \nu$ background but much more so in the description of 
the $B\to X_u \ell \nu$ signal.  

This paper is organised as follows. In Section~\ref{sec:setup}, we discuss the generation of the 
MC event samples used in our analysis, both with \Evtgen and \Sherpa, and show selected
distributions before and after an in-house detector simulation.  
In Section~\ref{sec:BDTvsNN}, we present the input features of our ML analysis and compare the performance of BDTs and NNs for different levels of input variables. 
While ML techniques have great potential in extending the fiducial regions of experimental
analyses, it is also vital to  understand their limitations. Therefore, 
in Section~\ref{sec:inclusivity}, we study the inclusivity of different ML approaches 
and their dependence on the MC generator used to produce the training data. 
Finally, we conclude in Section~\ref{sec:conclusion}.

\section{Event generation}
\label{sec:setup}

Our analysis aims at distinguishing $B\to X_u \ell \nu$ signal events from the
$\sim 50$ times larger background induced by the CKM-favoured $B\to X_c \ell \nu$ process. 
Other background contributions from continuum and combinatorial backgrounds are neglected. 
The training and test samples of the signal and background events for our ML analyses are produced using MC event generators. In this section we explain our simulation
set-up and explore characteristics of the signal and background before and after a detector simulation.  We also compare MC samples produced with the default generator for $B$-physics
analyses, \Evtgen-v01.07.00~\cite{Lange:2001uf}, with those from \Sherpa-v2.2.8~\cite{Gleisberg:2008ta}.

\subsection{Monte Carlo samples and event selection}  
Our event samples are generated at SuperKEKB/Belle~II beam energies of $4\,\gev$ and $7\,\gev$  or, equivalently, an $\Upsilon(4S)$ resonance with a four-momentum  of $p_{\Upsilon(4S)} = (11,0,0,3) \, \gev$. 

For the \Evtgen sample, we generate signal and background
events with the default run card. For the $B\to X_u \ell \nu$ signal we use the 
built-in hybrid model for combining resonant and non-resonant modes, with the default input values $m_b = 4.8 \,\gev$ for the $b$-quark mass, $a = 1.29$ for the Fermi motion parameter and 
$\alpha_s(m_b) = 0.22$ for the strong coupling at the $b$-quark mass~\footnote{For the resonant modes the following branching ratios for $B^0$ and $B^\pm$ are assumed: $\text{BR}(B^0 \to \pi^\pm \ell^\mp \nu)  = 1.5 \times 10^{-4}$, 
$\text{BR}(B^0 \to \rho^\pm \ell^\mp \nu)  = 2.94 \times 10^{-4}$; 
$\text{BR}(B^\pm \to \pi^0 \ell^\pm \nu)  = 0.78 \times 10^{-4}$, 
$\text{BR}(B^\pm \to \eta \ell^\pm \nu)  = 0.39 \times 10^{-4}$, 
$\text{BR}(B^\pm \to \rho^0 \ell^\pm \nu)  = 1.58\times 10^{-4}$, 
$\text{BR}(B^\pm \to \omega \ell^\pm \nu)  = 1.19 \times 10^{-4}$, 
$\text{BR}(B^\pm \to \eta' \ell^\pm \nu ) = 0.23 \times 10^{-4}$.}. 
The fragmentation of the $X_u$ system into final-state hadrons is performed by \Pythia~\cite{Sjostrand:2007gs,Sjostrand:2014zea}, and final state QED radiation is performed by \Photos~\cite{Barberio:1990ms,Barberio:1993qi}.

In the \Sherpa simulations, we make use of the standard run card for $B$-hadron pair production on the $\Upsilon(4S)$~pole and use the \Sherpa default settings for fragmentation. 

In both cases, our baseline event selection process is based on Ref.~\cite{Urquijo:2009tp}. We select events with one fully hadronically decaying $B$~meson on the tagging side ($B_\text{tag}$), and require the other $B$~meson on the signal side ($B_\text{sig}$) to decay semileptonically to an electron or muon with $p_\ell^{*}>1.0 \,\gev$, where $p_\ell^{*}$ is the magnitude of the electron or muon momentum in the $B$-meson rest frame.

\subsection{Detector effects}

\begin{figure}[thb]
  \centering
    \includegraphics[height=4.5cm]{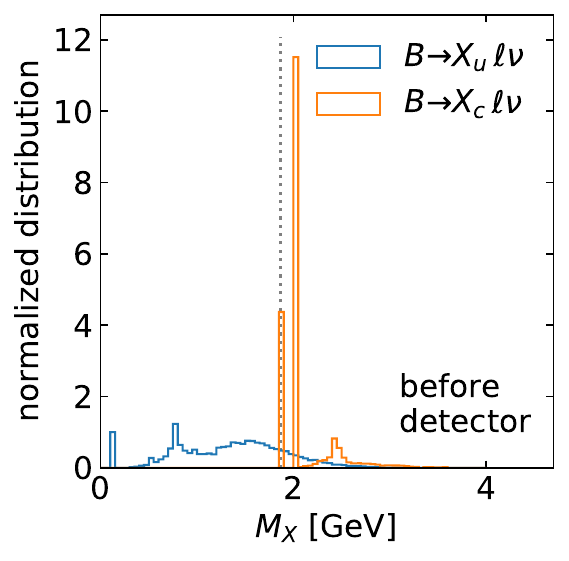}
    \;
    \includegraphics[height=4.5cm]{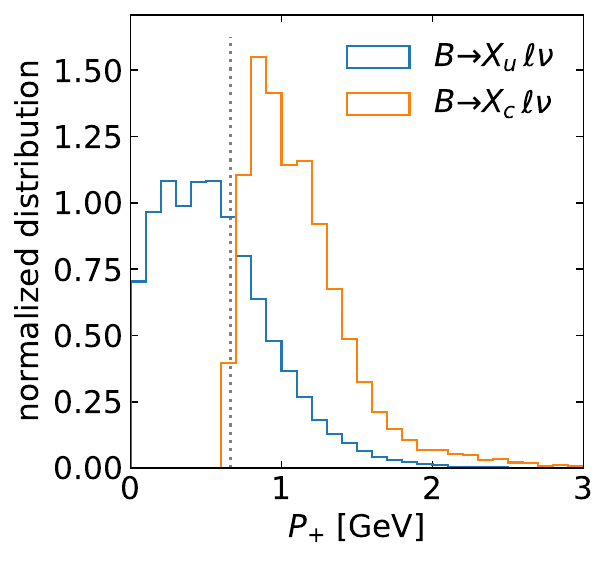} 
    \;
    \includegraphics[height=4.5cm]{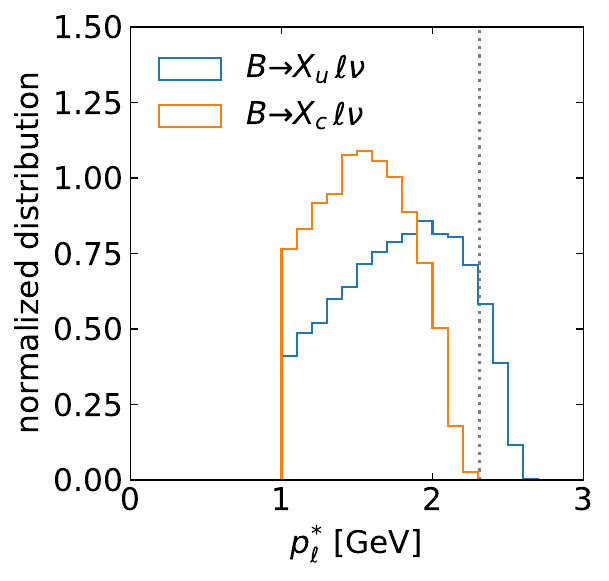}\\
    \includegraphics[height=4.5cm]{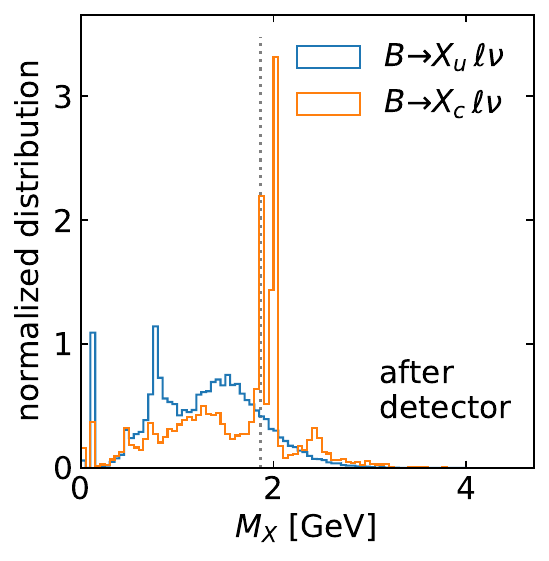} 
    \;
    \includegraphics[height=4.5cm]{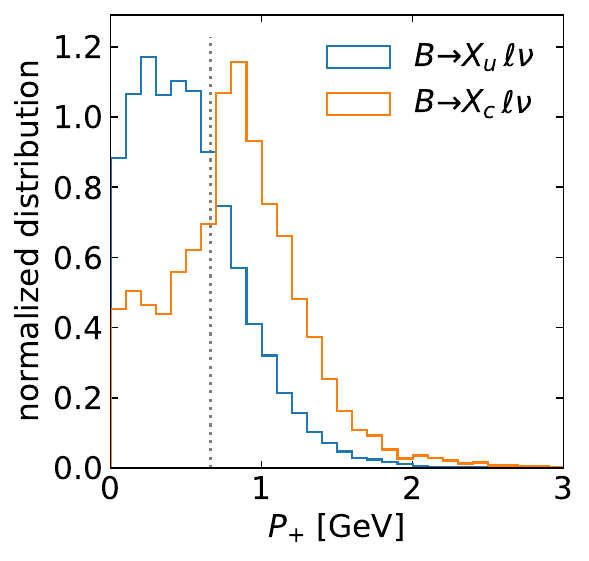} 
    \;
    \includegraphics[height=4.5cm]{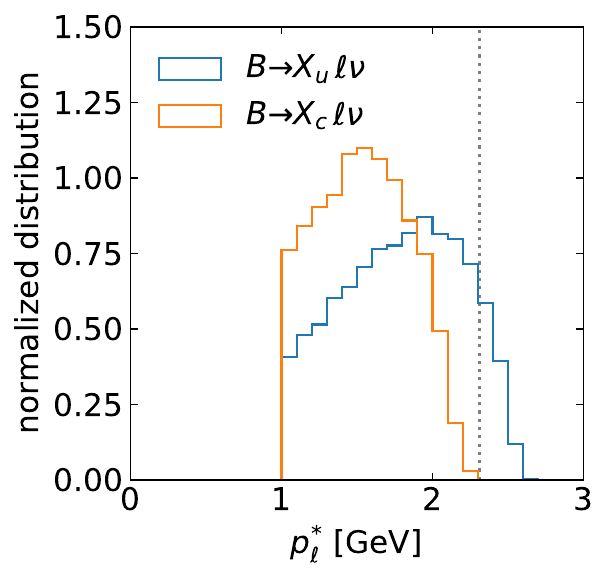}
  \caption{\Evtgen hadronic mass distribution $M_X$, energy-momentum difference $P_+$ and 
  lepton momentum in $B$-meson rest frame $p_\ell^*$
  before (top) and after detector simulation (bottom). The gray lines highlight the boundaries of the theoretically background-free regions. }
  \label{fig:mass_detector_yes_no}
\end{figure}

In order to mimic detector effects, we pass our MC data through an in-house
detector simulation described in Appendix~\ref{app:detector}. In that appendix we also
show some validation plots comparing our MC samples with those produced by 
the Belle collaboration (see  Figure~\ref{fig:validation1}). 

Our detector simulation includes detector efficiencies and mistagging for particles on the signal side; it does not take into  account that decay products from the tag side can be incorrectly assigned as signal-side particles. 
While this in-house detector simulation is too simplified to create completely
realistic event samples, it does show reasonable agreement with MC results from the
Belle collaboration, and can be considered sufficient for the purpose of the qualitative studies performed in this paper. 

In Figure~\ref{fig:mass_detector_yes_no}, we show normalized distributions 
of signal and background events in the \Evtgen MC sample before and after detector simulation 
for three kinematic variables: the hadronic invariant mass $M_X$, the energy-momentum difference $P_+$, and the lepton momentum in the $B$-meson rest-frame $p_\ell^*$.
The distributions of $M_X$ and $P_+$, which are based on multiple final-state particles and are therefore subject to a cumulative effect from detector inefficiencies and mistagging,
are clearly strongly affected by detector effects. 
In the low-$M_X$ and low-$P_+$ regions, detector effects cause the charm background to populate even the theoretically inaccessible phase-space regions $M_X <m_D$  and $P_+<m_D^2/m_B$.
The lepton momentum, on the other hand, can be determined quite precisely and detector effects have only a marginal effect.\footnote{This would also be the case in a more realistic simulation,
as long as the four-momentum of the tag-side $B$~meson, which determines the boost to the signal $B$-meson rest frame, is well reconstructed.} These plots make clear that to achieve an efficient separation of  signal and background after detector effects, kinematic cuts on their own are insufficient.  We will list a full set of distinguishing features of the signal used in our ML analysis
in Section~\ref{sec:features}.

\subsection{\Evtgen vs. \Sherpa}
\label{sec:EvtGenVsSherpa}

\begin{figure}[thb]
  \centering
     \includegraphics[width=0.3\textwidth]{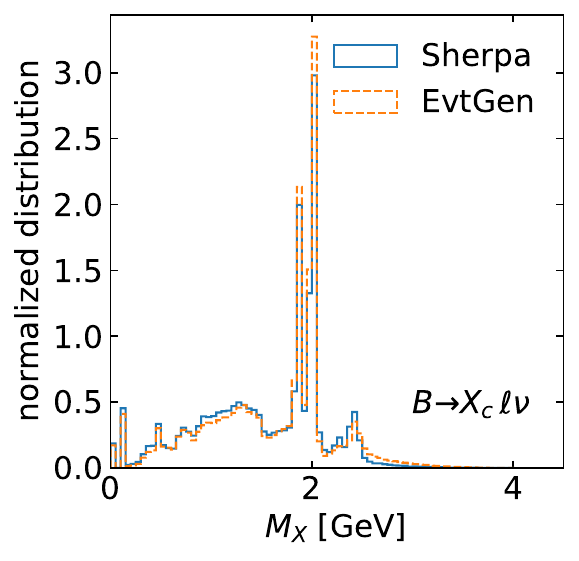}
     \;
     \includegraphics[width=0.3\textwidth]{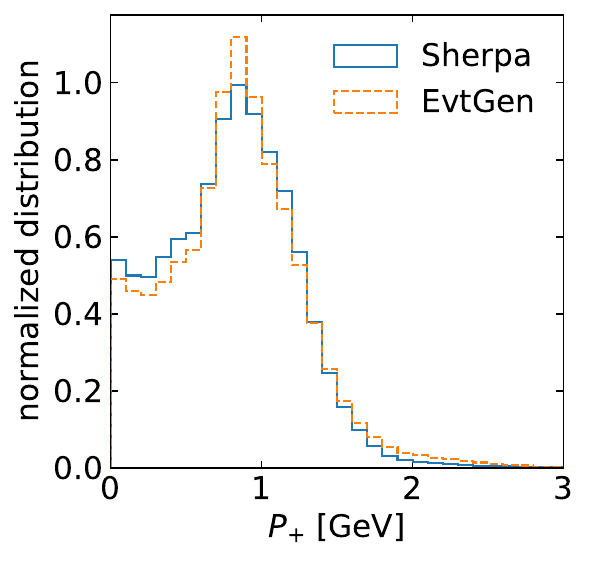}
     \;
     \includegraphics[width=0.3\textwidth]{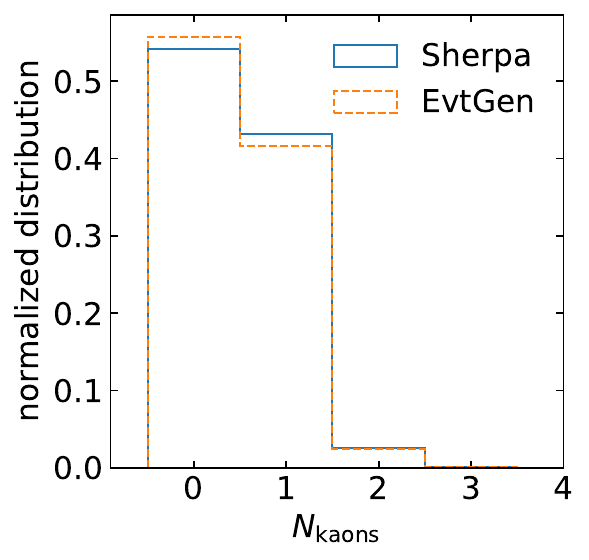} 
  \caption{High-level features of $B\to X_c\ell\nu$ events generated with \Evtgen and \Sherpa.}
  \label{fig:sherpa_vs_evtgen_high_level_features_bkg}
\end{figure}

While \Evtgen and \Sherpa follow the same general principle in modelling resonant
contributions, they differ in the treatment of the non-resonant modes.  In this
section we highlight the effects of these modelling choices on distributions 
of the signal and background.  

In Figure~\ref{fig:sherpa_vs_evtgen_high_level_features_bkg}, we compare distributions for the $B\to X_c\ell\nu$ background. In addition to the kinematic features $M_X$ and $P_+$
we also show the number of charged kaons $N_{\rm kaons}$ in the event.
Given that inclusive semileptonic decays into charm are nearly saturated by a small number of
resonant contributions, it is not surprising that the \Evtgen and \Sherpa results show a close agreement.  Minor differences, for instance the number of kaons,  are caused by small discrepancies in the assumed branching ratios for high-mass $X_c$ resonances as well as by the different hadronization modelling in \Pythia and \Sherpa.

\begin{figure}[thb]
  \centering
    \includegraphics[width=0.325\textwidth]{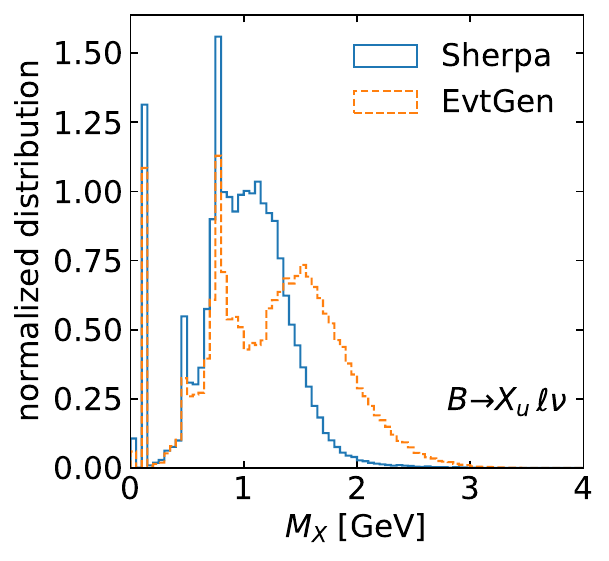}
    \,
    \includegraphics[width=0.304\textwidth]{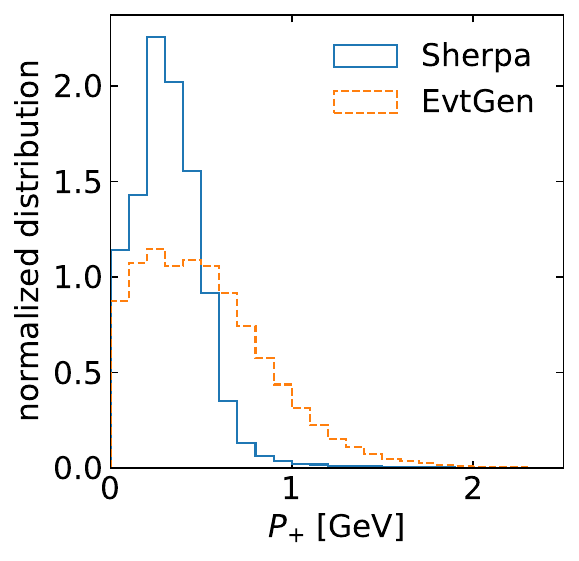}
    \,
    \includegraphics[width=0.309\textwidth]{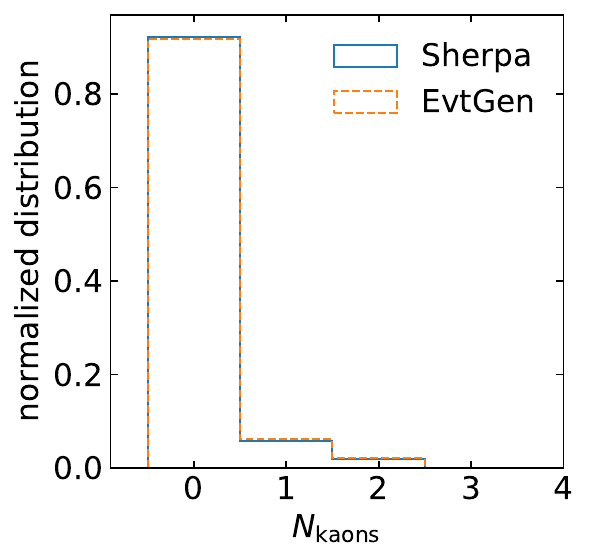}  \\
        \includegraphics[width=0.32\textwidth]{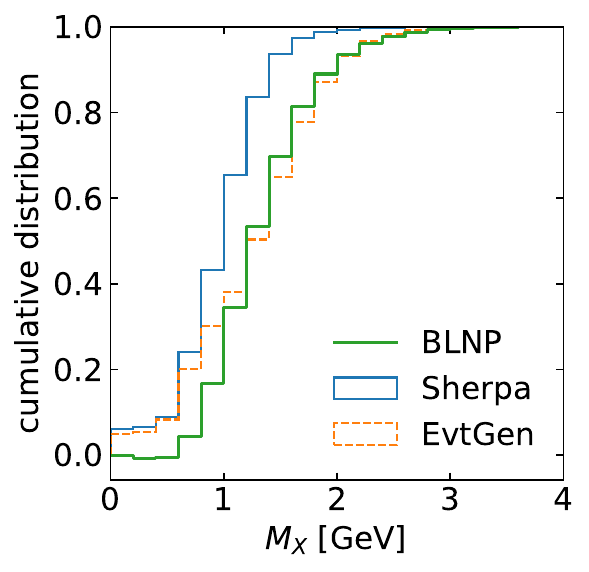} \,
    \includegraphics[width=0.32\textwidth]{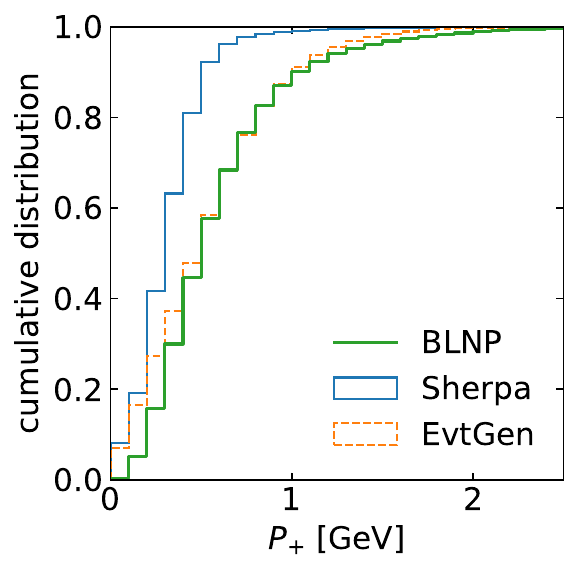} \,
    \includegraphics[width=0.32\textwidth]{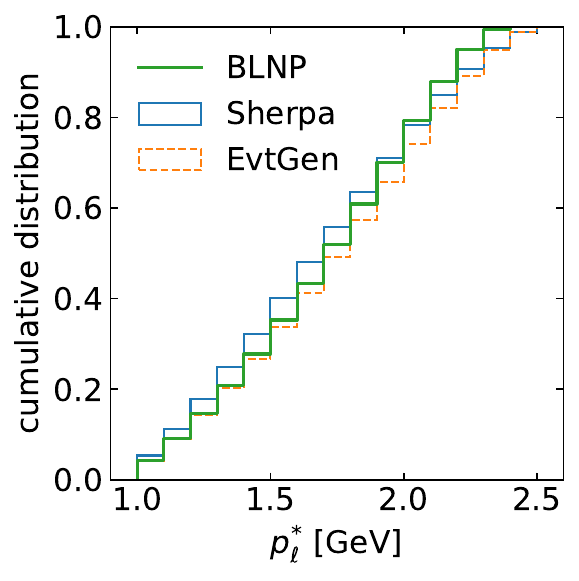}
  \caption{Upper panel: Comparison of \Evtgen and \Sherpa high-level features for $B\to X_u \ell \nu$ signal events.
  Lower panel: Cumulative sum of the differential distributions $M_X$, $P_+$ and $p_{\ell}^*$ in \Evtgen and \Sherpa, compared to BLNP prediction.}
  \label{fig:sherpa_vs_evtgen_high_level_features}
\end{figure}

The analogous distributions for the $B\to X_u\ell \nu$ signal are shown in the upper panel of Figure~\ref{fig:sherpa_vs_evtgen_high_level_features}.
There are clear differences between the \Evtgen and \Sherpa distributions of kinematic features such
as the $M_X$ distribution, which are caused by the different treatment of the non-resonant modes.
In \Evtgen, the built-in hybrid model describes the non-resonant decay modes at leading order in the heavy-quark expansion using the DeFazio-Neubert~(DFN) model~\cite{DeFazio:1999ptt}, including a non-perturbative shape function to describe the Fermi motion of the $b$~quark inside the $B$~meson.  The non-resonant contribution is modelled such that the $M_X$ distribution for 
the sum of the resonant and non-resonant contributions matches the distribution predicted by the DFN model.   This is achieved through a bin-by-bin reweighting of the non-resonant modes.
 
In \Sherpa  the non-resonant signal decay modes are modelled by parton showering and hadronizing
the leading-order partonic decay.  
Non-perturbative shape-function effects characterising the low-$M_X$ region
are not taken into account, and no reweighting of the events is performed to match state-of-the-art theory calculations.  

Comparing these two different approaches for the signal modelling in Figure~\ref{fig:sherpa_vs_evtgen_high_level_features}, we find that, 
on the one hand, the \Evtgen results have a non-physical bump in the $1.5\,\gev$ region of the $M_X$ distribution,
which is an artefact of the bin-by-bin reweighting to match the DFN results. The
\Sherpa distributions do not share this characteristic, since the non-resonant events are instead
obtained by excluding resonant events from the parton shower.  On the other hand, the current
implementation of the \Sherpa parton shower model also produces a smaller proportion of the non-resonant signal contribution and generates fewer events in the 
high-$M_X$ and -$P_+$ regions compared to \Evtgen, which is precisely the region 
where the inclusive QCD predictions should be reliable. 
We further highlight this in the lower panel of Figure~\ref{fig:sherpa_vs_evtgen_high_level_features}, where we compare the state-of-the-art OPE results from the BLNP approach~\cite{Lange:2005yw} with \Evtgen and \Sherpa results at the level of cumulative 
distributions.  Overall, the agreement of the \Evtgen-generated distributions with the BLNP predictions is stronger, which is not surprising since the underlying inclusive is the OPE-based
DFN result.  

Clearly, the $B\to X_u\ell \nu$ modelling in \Sherpa  
needs a more sophisticated matching of the non-resonant, parton shower 
contributions with (shape-function) OPE results before being used in $\vub$ extractions
by experiments.  For this reason, we use \Evtgen in the following section when studying
the performance of ML-based classifiers, in spite of its own deficiencies in the low and 
intermediate invariant mass regions. However, for the purposes of the paper, the present
situation allows us to study an interesting question: 
how do ML approaches to $\vub$ extractions perform when the training and testing 
data are substantially different?
This is the subject of Section~\ref{sec:inclusivity}.

\section{BDTs vs Deep Neural Networks}
\label{sec:BDTvsNN}

In this section we give a systematic analysis of signal vs.\ background 
event classification using BDTs and deep neural networks.
We use Bayesian neural networks~(BNNs), which have been argued to deliver stable results and avoid overfitting~\cite{Bollweg:2019skg}. The details of the architecture for the BDTs and 
NNs used in our study can be found in Appendix~\ref{app:MVA_setup}, along with a breakdown of data used in the training  and testing procedure.  We describe the input features to the ML
algorithms in Section~\ref{sec:features},  metrics used in evaluating their performance in 
Section~\ref{sec:metrics}, and then move on to the results in Section~\ref{sec:Results}.  
Throughout this section we use \Evtgen to generate the training and testing samples.

\subsection{Input features}
\label{sec:features}

The features used in our multivariate analysis break into two sets.
One is based on physical high-level features such as invariant masses and the 
number of final-state particles of a specific type, e.g.\ the number of kaons or slow pions, 
and the other is based on low-level features, i.e.\ single particle properties.  In particular,
the low and high-level features are:
\begin{itemize}
	\item \textbf{low level}
	\begin{equation}
 p_{B_\text{tag}} ,\, 
Q_{B_\text{tag}}, \, 
 p_i , \, 
 \text{ID}_i , \, 
Q_i 
     \quad i \in \text{top 10 most energetic particles.}
\label{eq:features_low}    
\end{equation}
	\item \textbf{high level}
	\begin{equation}
\begin{split}
  q^2, \quad  M_X ,\quad
    &P_+ ,\quad
    p_\ell^{*} ,\quad
    N_\ell ,\quad
    N_{K^\pm} ,\quad
    N_{K^{0}} ,\quad
    N_\text{hadron} , \quad
    M_\text{miss}^2 , \quad
    Q_\text{tot} , \\
    &N_{\pi^0_\text{slow}}, \quad
    N_{\pi^\pm_\text{slow}} , \quad
    M_{\text{miss},\, D^*}^2(\pi^0_\text{slow}) , \quad
    M_{\text{miss},\, D^*}^2(\pi^\pm_\text{slow}) .
\end{split}
\label{eq:features_high}
\end{equation}
\end{itemize}
The low-level features include, first off, the four-momentum $p_{B_\text{tag}}$ and charge 
$Q_{B_\text{tag}}$ of the tagged $B$~meson. In addition, we  pick out the 10 most 
energetic (as measured in the lab frame) detected final-state particles, label
them  with an index $i=1,\dots 10$, and use as features the lab frame four-momenta $p_i$, 
the charge $Q_i$ and the identity $\text{ID}_i$ of these particles. Events with less than 10 detected
final-state particles have the corresponding particle features filled in with zeros. 

\begin{figure}[t]
  \centering
    \includegraphics[height=4.4cm]{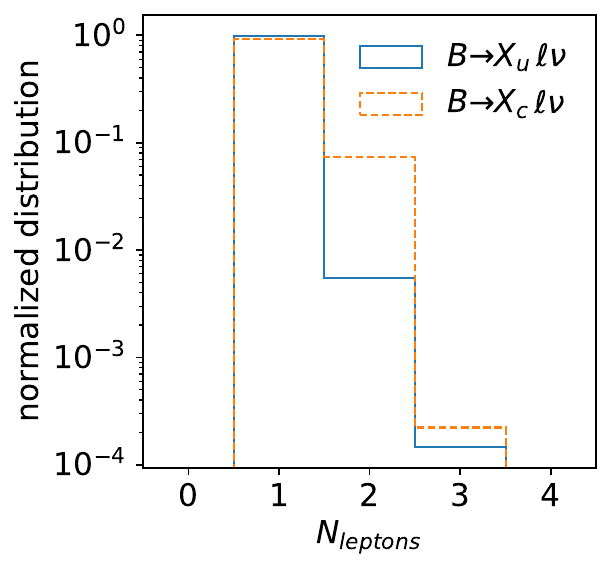}
    \;
    \includegraphics[height=4.4cm]{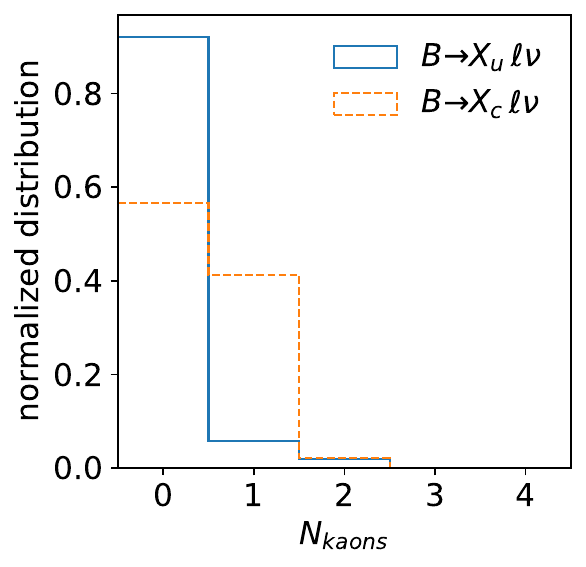}
    \;
     \includegraphics[height=4.4cm]{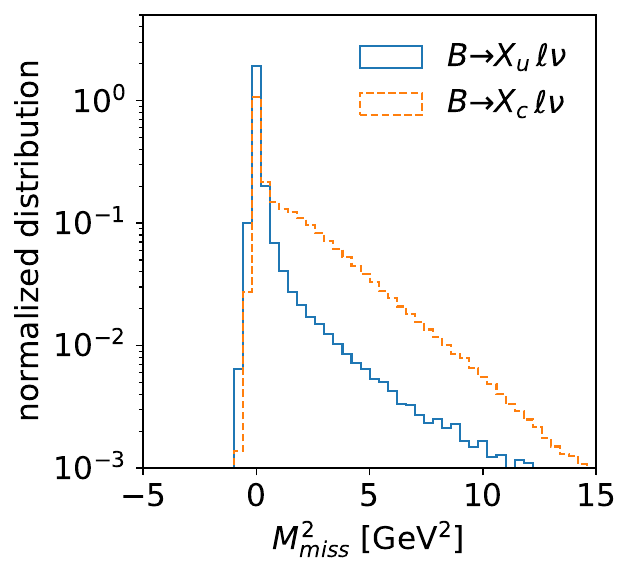}
  \caption{High-level features of the \Evtgen sample. 
  Number of leptons $N_\ell$ (left), number of kaons~$N_\text{kaons}$ (middle) and missing mass squared~$M_\text{miss}^2$ (right). Notice the logarithmic scale for some of the distributions. 
  }
  \label{fig:high_level_features}
\end{figure}

The high-level features are defined as follows. The four-momentum transfer squared is $q^2 = (p_B - p_X)^2$. $N_\ell$ denotes the number of leptons, which can only be greater than one if the secondary leptons have momenta smaller than $1\,\gev$.
Since the $B \to X_u \ell \nu$ signal is very unlikely to contain secondary leptons, this feature can be used to suppress the background, see the left panel of Figure~\ref{fig:high_level_features}.
$N_{K^\pm}$ and $N_{K^{0}}$ denote the number of charged and neutral kaons, respectively, where neutral kaons $K^0_S$ are reconstructed from charged pions with an invariant mass in the
range $m_{\pi^+ \pi^-} \in [0.490, \, 0.505]\,\gev$.
Kaons are frequently produced in $D$-meson decays and their presence hence indicates a $B \to X_c \ell \nu$ background event, see the central panel of Figure~\ref{fig:high_level_features}.
The number of final-state particles resulting from the hadron decay $N_\text{hadron}$ is typically larger for hadrons with a higher mass such as the background $D$~mesons. 
The missing mass squared $M_\text{miss}^2$, 
defined as the square of the missing momentum 
$p_\text{miss} = p_\text{sig}-p_X - p_\ell$, where $p_\text{sig} = p_{\Upsilon(4S)} - p_{B \text{tag}}$ is the reconstructed momentum of the signal-side hadron, would always be compatible with zero without detector effects. For background events, which as discussed above have a higher final-state particle multiplicity, the probability of misidentifying a final-state particle is higher resulting in positive values of the missing mass squared, see the right panel of Figure~\ref{fig:high_level_features}.
The total charge $Q_\text{tot}$ of all particles in the event, on both the signal and the tag side, is also subject to detector effects. It will only be non-zero for events where charged particles have been missed, which happens more often for the background events due to their larger final-state particle multiplicity.  
Slow pions, i.e.\  pions with momentum $|p_\pi|< 220\,\mev$, can originate from $D^* \to D \pi$ transitions and hence appear more often for the $B \to X_c \ell \nu$ background.
We therefore include the number of neutral and charged slow pions, $N_{\pi^0_\text{slow}}$ and $N_{\pi^\pm_\text{slow}}$, in our high-level feature set.
To test the compatibility of the slow pion with a $D^* \to D \pi$ transition, we further define $M_{\text{miss},\, D^*}^2 = (p_\text{sig}-p_{D^*}-p_\ell)^2$, where $p_{D^*} = (E_{D^*}, \vec{p}_{D^*})$ with $E_{D^*}= \frac{m_{D^*}}{m_{D^*}-m_D} E_\pi $ and $\vec{p}_{D^*} = \vec{p}_\pi \frac{\sqrt{E_{D^*}^2-m_{D^*}^2}}{|\vec{p}_\pi|}$. In this we have explicitly assumed that the slow pion direction is strongly correlated with the $D^*$ direction. The quantity $M_{\text{miss},\, D^*}^2$ will more likely be peaked at zero for true $D^* \to D \pi$ transitions.
Distributions in the high-level input features not shown in Figure~\ref{fig:high_level_features} are displayed in Appendix~\ref{app:morePlots} in Figure~\ref{fig:more_high_level_features}.

We have chosen this set of high-level features to mimic the feature selection in the 
BDT analyses performed by Belle in Refs.~\cite{Urquijo:2009tp, Cao:2021xqf}.
Some differences with respect to the sets used in those papers arise, because we do not have access to all experimental features in our simplified detector simulation, for instance features related to the quality of the signal reconstruction.

\subsection{Metrics}
\label{sec:metrics}

Before we compare the performance of different ML approaches and input feature set-ups, let us briefly introduce some notation for the ML output and review metrics used
to quantify performance.

Our binary classifiers take as input the multidimensional features of an event, and return
a classifier output which is a single number, $\zeta \in [0, \, 1]$. 
Events with classifier output $\zeta \sim 1$ are likely to be signal  while events with $\zeta \sim 0$ are likely to be background.  We define our signal (fiducial) region through a cut on the 
classifier output.  All events with $\zeta > \zeta_\text{cut}$ are classified as signal events. 
Events which are correctly classified as signal events are denoted true positive (TP) events, while background events which are incorrectly classified as signal events
are denoted false positive (FP) events. 

Standard performance metrics in ML are the receiver operating characteristic (ROC) curve, i.e. the 
\textit{true positive rate} (TPR, signal acceptance) as a function of the \textit{false positive rate} (FPR, background acceptance), and the corresponding area-under-curve~(AUC), the integral of the ROC curve.
It is also customary to plot the inverse of the FPR as a function of the TPR. 
A quantity which is often used as a metric in particle physics is the statistical 
significance~$\sigma$, defined as 
\begin{equation}
    \sigma  = \frac{\TP}{\sqrt{\TP+\FP}} = \frac{S}{\sqrt{S + B}} \, .
    \label{eq:significance}
\end{equation}
where in the second equation we have used $S$ and $B$ to denote the number of signal and background events in the signal region
to bring the expression into a more familiar form. 
To remove the dependence on the data sample size from the significance, we make use of the significance improvement~$\hat{\sigma}$,
i.e.~the significance normalized to its value at the baseline selection
\begin{equation}
    \hat{\sigma} = \frac{\sigma }{\sigma_\text{baseline} } 
    \label{eq:SIC}
\end{equation}
A significance improvement greater than one signals a performance increase. 
Plotting the significance improvement as a functions of the true positive rate defines the significance improvement characteristic~(SIC) curve~\cite{Gallicchio:2010dq}. 

\subsection{BDT and BNN performance on different levels of input features}
\label{sec:Results}
\begin{figure}[thb]
  \centering
    \includegraphics[width=0.48\textwidth]{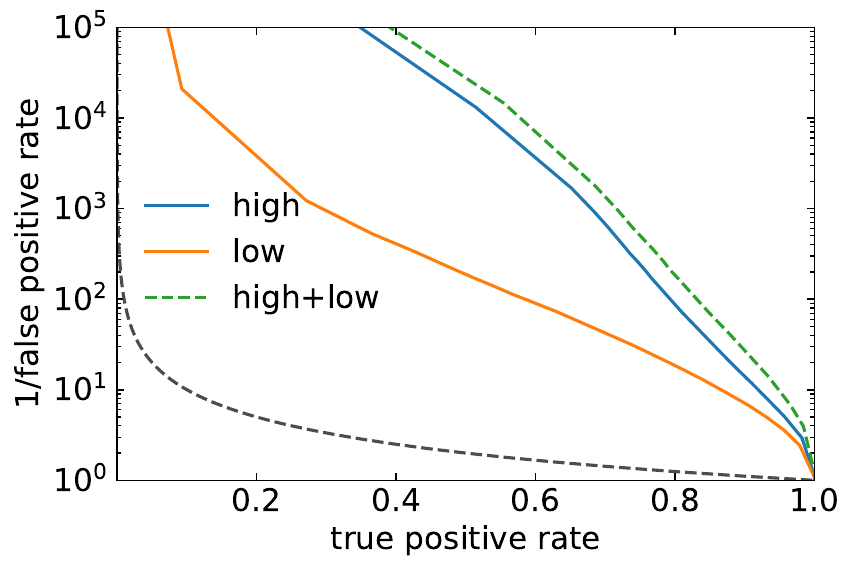}
    \;
     \includegraphics[width=0.48\textwidth]{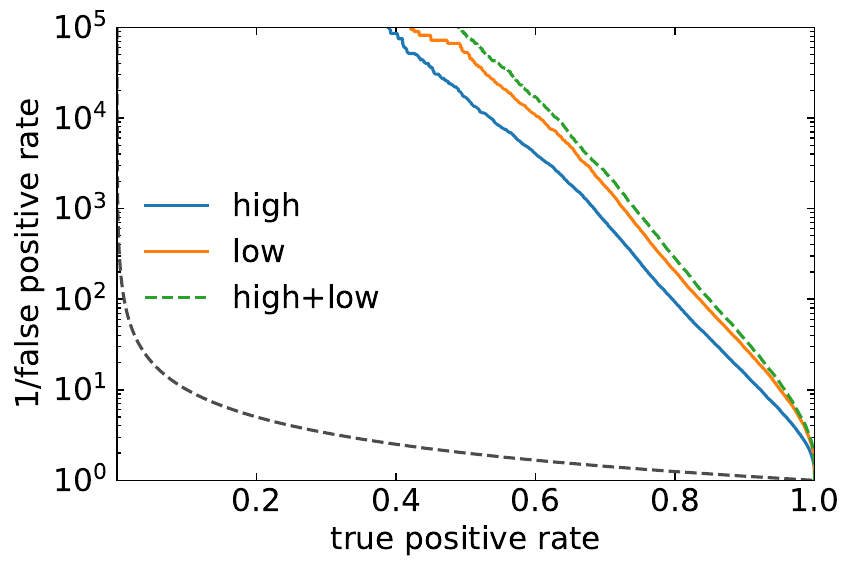}
     \\
     \includegraphics[width=0.48\textwidth]{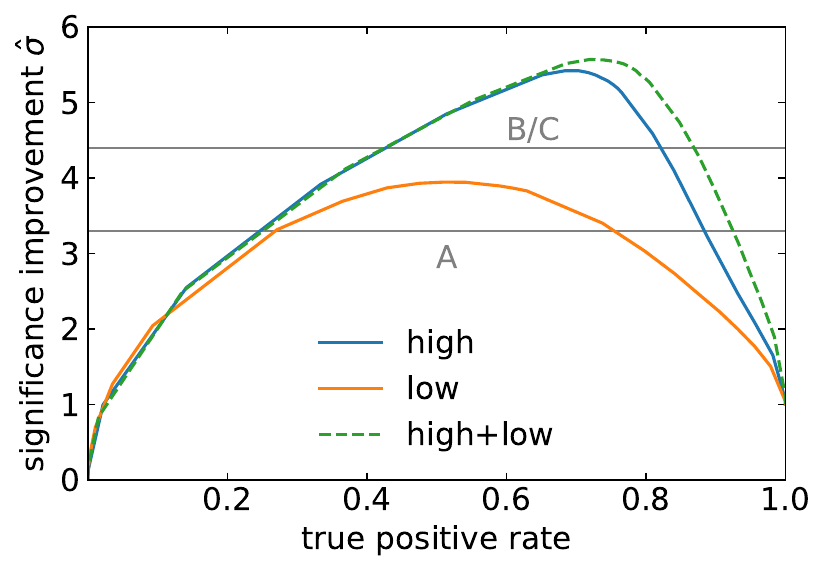}
    \;
     \includegraphics[width=0.48\textwidth]{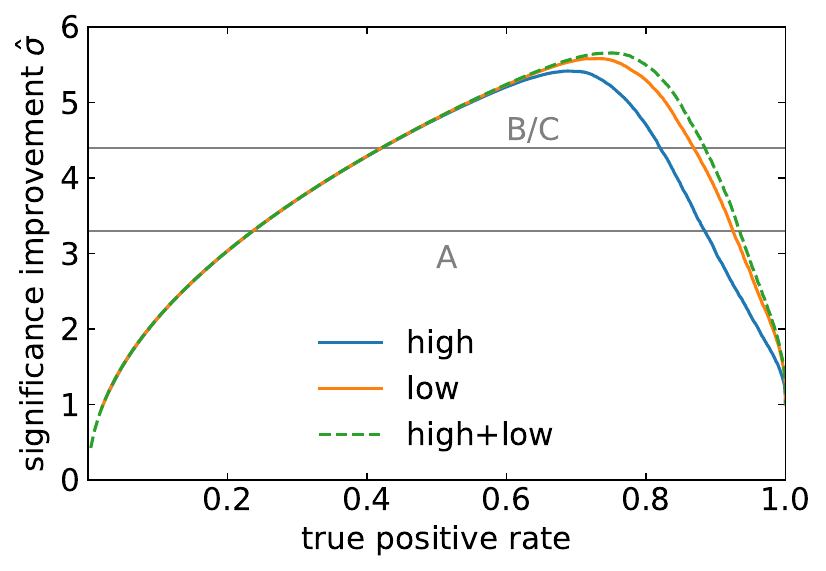}

  \caption{ROC (top) and SIC curves (bottom) for BDT (left) and BNN (right) for different
  levels of input features, trained and tested on \Evtgen data with a physical ratio of signal-to-background events in the test set. The dashed lines in the upper panel are ROC curves 
for the case of no separation. As a reference, the gray lines in the bottom panel
show the significance improvement from the three cut-and-count scenarios in Eq.~\eqref{eq:ref_cut_and_count}. A: $M_X < m_D$, B: $M_X < 1.5 \,\gev$, C: $P_+< m_D^2/m_B$.}
  \label{fig:roc_high_low}
\end{figure}

We first contrast the performance of the BDT and BNN on signal vs.\ background classification
using different levels of input features. We consider three scenarios: 
\begin{enumerate}[(i)]
    \item using only the 
low-level features in Eq.~\eqref{eq:features_low}
    \item using only the high-level features in Eq.~\eqref{eq:features_high}
    \item using a combination of these low- and high-level features.
\end{enumerate}
The ROC and SIC curves for the BDT and BNN analyses using these input feature scenarios are shown in 
Figure~\ref{fig:roc_high_low}. 

As expected, the BDT performs well on high-level input features, the most used features being the number of kaons, number of leptons, the hadronic invariant mass $M_X$, hadron multiplicity and the missing mass squared $M_\text{miss}^2$.
However, it performs poorly when trained only with low-level features, indicating that it cannot 
use them to construct additional non-linear features such as invariant masses.
Using a combination of low- and high-level features slightly improves the BDT performance
compared to high-level only. We have explicitly checked that this performance increase results almost entirely from adding the particle energies. The particle three-momenta, on the other hand, do not seem to contain additional usable information for the BDT. 

For the BNN the situation is very different.  It performs slightly better when trained only on low-level
features than it does when trained only on high-level features. This indicates that, as expected,
it is able to learn new and efficient discriminating features from the low-level inputs. 
Training on a combination of low plus high-level inputs very marginally improves its
performance compared to low-level only (mainly due to the inclusion of $M_X$ as a feature), showing that the BNN has learned the most 
important high-level features on its own. 

The maximum of the SIC curves is reached for a cut on the classifier output of $\zeta_\text{cut} \approx 0.97$, which corresponds to a signal acceptance, or true positive rate $\TPR = \TP/(\TP+\text{FN})$, of approximately $75\,\%$. 
Explicitly, we find the following values for the maximum significance improvement and the AUC for a BDT or BNN trained and tested on a combination of high and low-level features from the
\Evtgen data:
\begin{align}
\begin{split}
   \AUC &= 0.981 , \quad \hat{\sigma} = 5.59 \; \quad \text{BDT} \\
   \AUC &= 0.986 , \quad \hat{\sigma} = 5.67 \; \quad \text{NN}\, . 
   \label{eq:best_result}
\end{split}
\end{align}
The AUC and $\hat{\sigma}$ for the BNN is only about 2\,\% better than BDT approach. 
Training on high-level features only puts the BNN on equal footing with the BDT -- in fact, we find that in that case they reach the exact same significance improvement, which is $\hat{\sigma}=5.42$.
The very small loss of performance compared to the Eq.~(\ref{eq:best_result}) indicates
that the high-level features are well chosen for a discrimination of signal and background, 
containing (almost) the full relevant information that the BNN can learn from the low-level features.     

It is interesting to contrast the significance improvements using the BDT and BNN with
those obtained from a typical cut-and-count analysis based on the cuts provided in Ref.~\cite{Lees:2011fv}.  With the minimal requirement of having exactly one lepton, a total charge of zero, a veto on kaons and a low missing mass squared,
\begin{equation}
N_\ell =1 , \quad 
Q_\text{tot}=0, \quad 
N_\text{kaons}=0, \quad 
M_\text{miss}^2 < 0.5 \, \gev^2 \, ,
\label{eq:cuts_cut_and_count}
\end{equation}%
we obtain a significance improvement of $\hat{\sigma} = 1.9$. If in addition to theses cuts we select a theoretically background-free region, we find\footnote{We consider the cut scenario $M_X< 1.5\,\gev$ in addition to $M_X < m_D$ to account for the fact that the background will dominantly populate the region slightly below $m_D$ due to detector effects, see Figure~\ref{fig:mass_detector_yes_no}.}
\begin{align}
\begin{split}
    \hat{\sigma}(M_X < m_D) = 3.3 , \quad \quad 
    \hat{\sigma}(M_X < 1.5 \,\gev) = 4.4 ,  \quad \quad 
    \hat{\sigma}(P_+< m_D^2/m_B) = 4.4 \, .
\end{split}
\label{eq:ref_cut_and_count}
\end{align}
Comparing the significance values Eq.~\eqref{eq:ref_cut_and_count} with those from the 
BDT and BNN analysis in Eq.~(\ref{eq:best_result}), we see that the ML approaches clearly 
outperform the cut-and-count analyses. In Appendix~\ref{app:increased_resolution}, we study the dependence of these results on the detector simulation.


\section{Inclusivity of ML approaches}
\label{sec:inclusivity}

A main motivation for the application of ML techniques to \vub determinations is to widen 
the experimentally accessible fiducial region to a level of inclusivity where the theoretically
clean, local OPE is unambiguously applicable. 
This amounts to two conditions on the measured $X_u$ final state:
first, that it is not subject to severe kinematic cuts (in which case the shape-function OPE would 
apply), and second, that it contains a sufficiently broad sample of exclusive hadronic 
final states in a given kinematic region (such that quark-gluon duality applies).
A concern in supervised ML approaches is that the classifiers will overuse either 
inclusive kinematic properties or IR unsafe hadron-level properties of 
the final state, thereby limiting the signal output to a restricted fiducial region
which is very sensitive to MC modelling, regardless of the inclusivity of the input events.

In this section we study the inclusivity of the signal acceptance in ML approaches to event
classification.  As the inclusivity depends crucially on the input features used in the 
ML classifier, we consider two scenarios:
\begin{itemize}
\item {\bf \NNone}: a NN using as input both the low and high-level features listed in Eq.~(\ref{eq:features_low}) and Eq.~(\ref{eq:features_high}), respectively.  
This is a more sophisticated implementation of the basic approach of Ref.~\cite{Urquijo:2009tp},
and its classification power was explored in Section~\ref{sec:Results}.
\item {\bf \NNtwo}: a NN using as input the high-level features listed in Eq.~(\ref{eq:features_high}), but {\it excluding} the kinematic features $M_X,\, P_+, \, q^2$ and $p_\ell^{*}$.  This is a proxy for 
the BDT used in the recent reanalysis of Belle data \cite{Cao:2021xqf}.   
\end{itemize}
In both cases the classifier threshold is chosen to maximize the significance of the accepted event set. Obviously \NNtwo,  which intentionally excludes discriminating kinematic features of the 
signal and background,  will not lead to the same signal purity as \NNone. In our analysis
\NNone reaches a signal-over-background ratio of $S/B \sim 13$, while for \NNtwo 
 $S/B \sim 0.3$ such that the background contribution is still dominant even
after event selection by the BNN. In this latter case  it is thus essential to perform a binned one- and two-dimensional likelihood analyses of the kinematic features of the signal and background after event selection by the \NNtwo,  as was done in Ref.~\cite{Cao:2021xqf}; this procedure can be 
useful for  \NNone as well, even though the $S/B$ ratio is much higher.

A main focus of our study is how changes of the testing and training data affect the inclusivity of the ML analyses. Testing and training the BNNs on differently modelled event sets provides a good test for overtraining and gives insight into how well the classifier might perform when applied to real-world events, which are not expected to show perfect  agreement with MC data.  The existing ML-based Belle analyses \cite{Urquijo:2009tp, Cao:2021xqf}  estimate uncertainties stemming from input data modelling by testing on samples produced with different parameter choices within the \Evtgen framework while fixing the ML configuration. Here we explore the alternative method of using a fundamentally different MC-event generation framework, namely \Sherpa. In this section we train all BNNs on \Evtgen and then study their classification properties on both \Sherpa and
\Evtgen data; in Appendix~\ref{sec:Sherpa_Training} we show equivalent results when the BNNs are trained instead on \Sherpa data.  All MC samples used in testing the BNNs, whether generated by \Sherpa or \Evtgen, contain the same ratio of signal to  background events after detector simulation.

We compare the inclusivity of the two BNN set-ups in two main ways.  In
Section~\ref{sec:signal_acceptance}, we study the inclusivity in kinematic phase
space, and in Section~\ref{sec:hadronization} we focus on inclusivity in the 
available hadronic final states.  In the latter section we also study sensitivity to changes of 
hadronization parameters within the \Evtgen framework.

\subsection{Inclusivity in kinematics}
\label{sec:signal_acceptance}

\begin{figure}[tbh]
\centering
	\includegraphics[height=5.5cm]{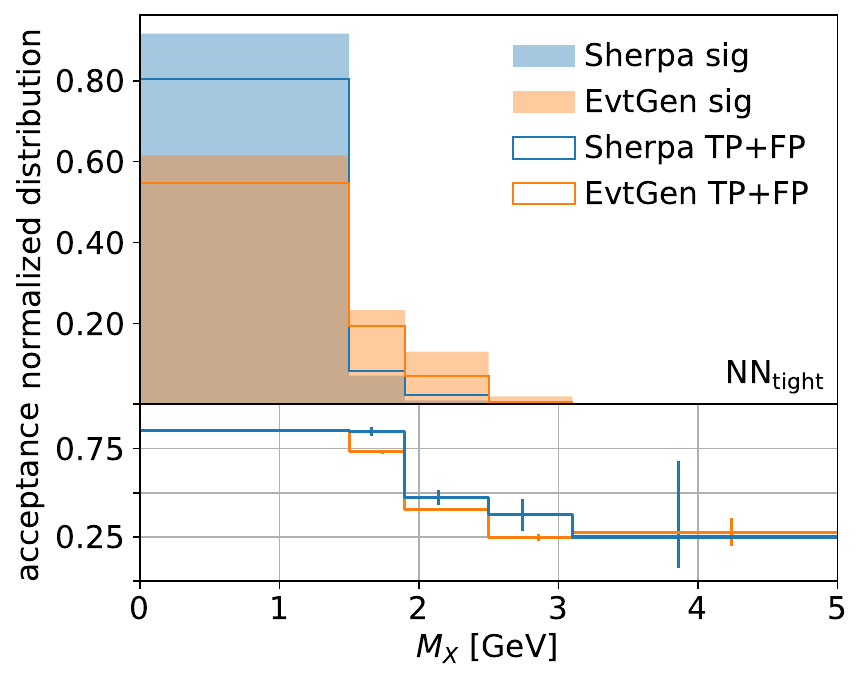} \quad 
\includegraphics[height=5.5cm]{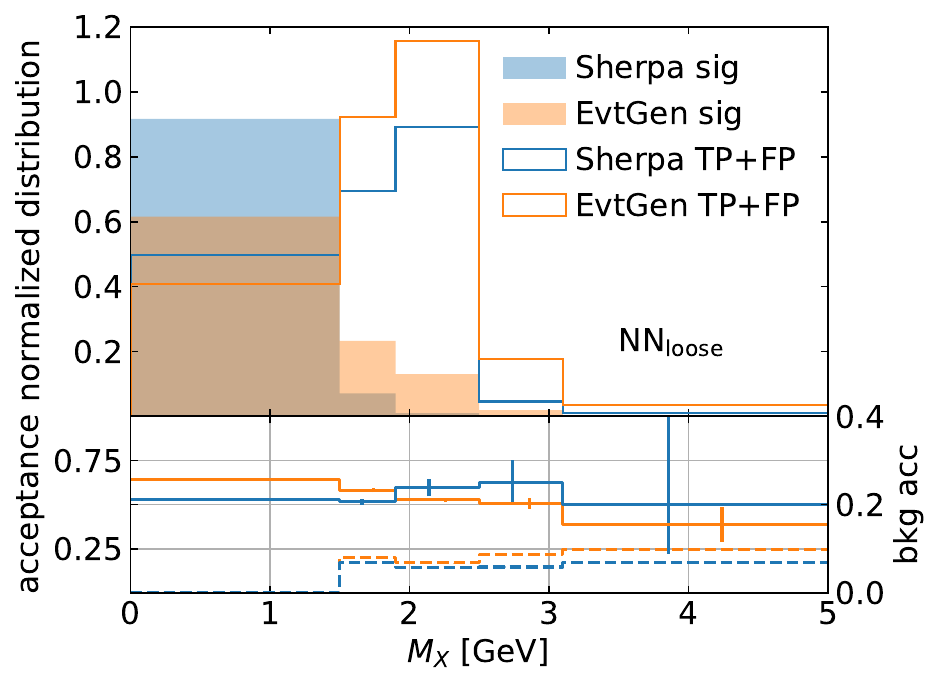}
\\[2mm]
\includegraphics[height=5.5cm]{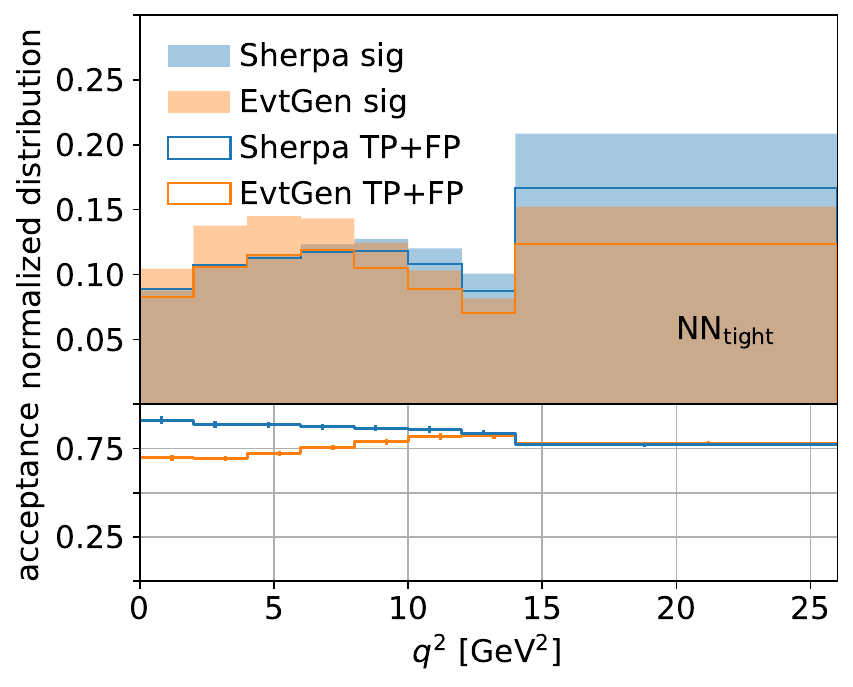}\quad
	\includegraphics[height=5.5cm]{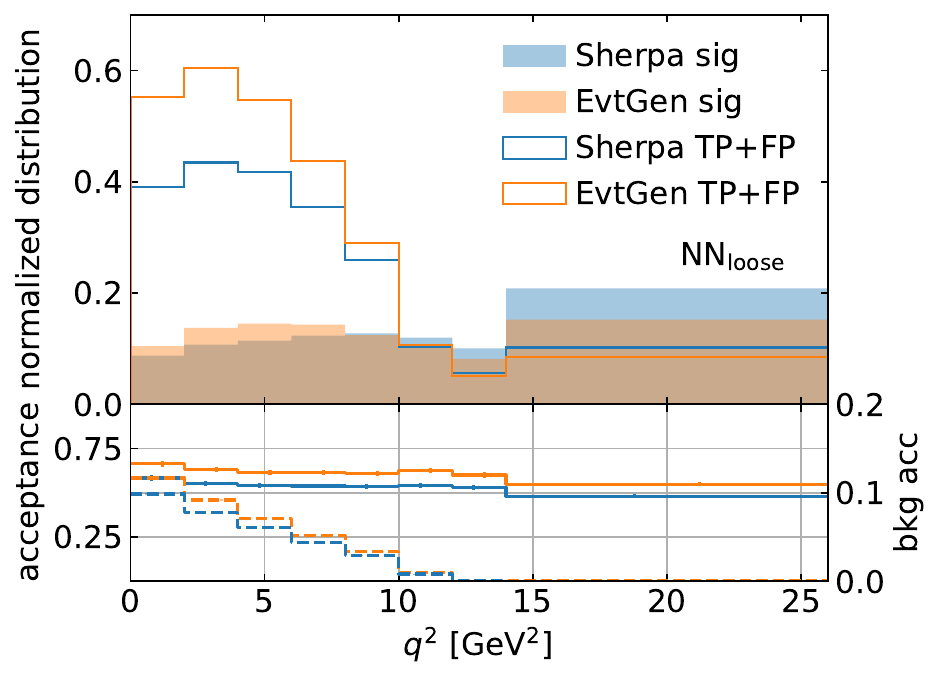}
  \\[2mm] 
   	\includegraphics[height=5.5cm]{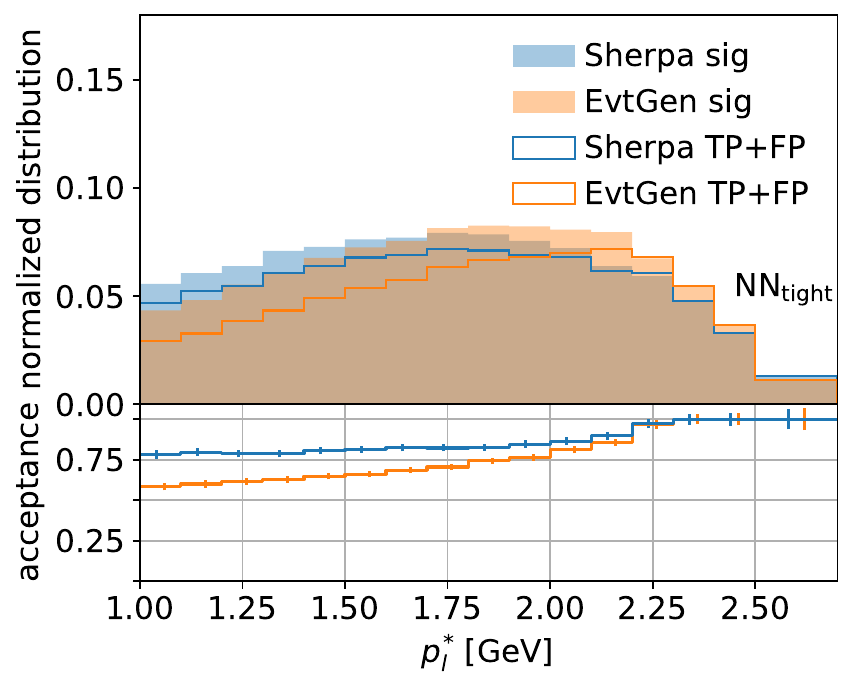}\quad
	\includegraphics[height=5.5cm]{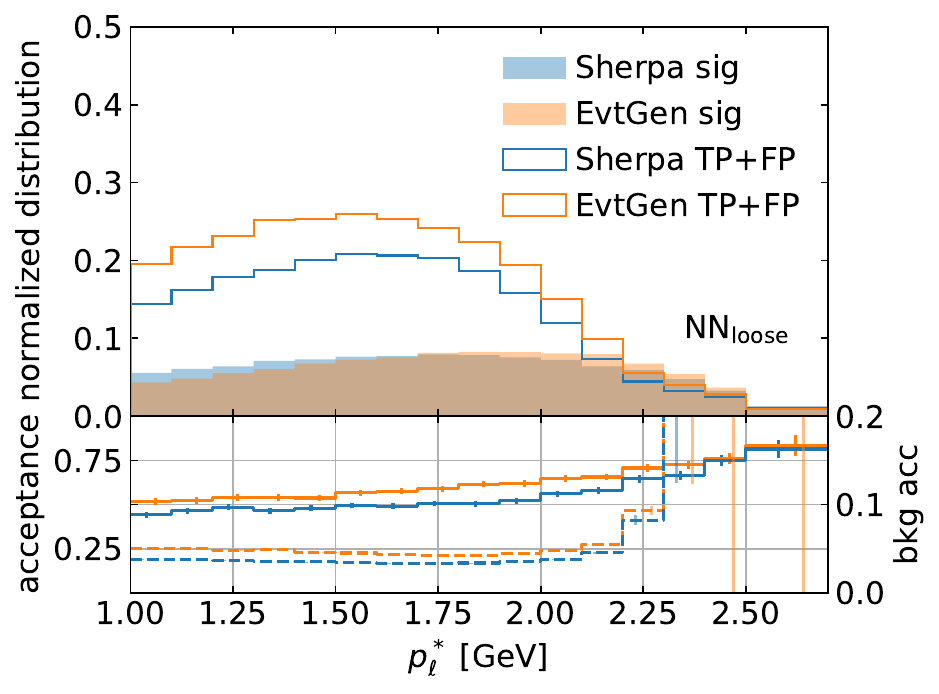}
    \\[2mm]
	\caption{Distributions and signal acceptance of \Sherpa and \Evtgen Monte Carlo 
	data as functions of $M_X$, $q^2$, 
	and $p_\ell^*$ for \NNone (left) and \NNtwo (right), trained on \Evtgen data. The distributions in the upper panels of each plot are normalized to the total number of signal events.
For \NNtwo the dashed lines in the lower panels show the background acceptance, using the scale for the $y$-axis displayed on the right. In the lower panels, error bars highlight the MC uncertainty on the acceptance. The error bars for the background uncertainty, which becomes visible at high $p_\ell^*$, uses a lighter shade. For bins with very low event numbers, we have used the tabulated uncertainties from Ref.~\cite{JAMES1980475}.}
	\label{fig:high_extra_acceptance}
\end{figure}
We illustrate the salient features of event selection by \NNone and \NNtwo as a function of 
$M_X$, $q^2$, and $p_\ell^*$  in Figure~\ref{fig:high_extra_acceptance}.
The binning of the kinematic variables matches that used in the fitting procedure of the 
recent $|V_{ub}|$ extractions in Ref.~\cite{Cao:2021xqf}:  
\begin{equation}
    \begin{split}
        M_X &= [0, \, 1.5, \, 1.9, \, 2.5, \, 3.1, \, 4.0] \, \gev  \, ,\\
         q^2 &= [0, \, 2, \, 4, \, 6, \, 8,\, 10,\, 12, \,14,\, 26] \, \gev^2 \, , \\
         p_\ell^* &= \mbox{15 equidist.~bins in } [ 1, \, 2.5] \, \gev \mbox{ \& } \,  [ 2.5, \, 2.7] \, \gev  \, .
    \end{split}
\end{equation}
In all cases, the bins are sufficiently wide that the results can be compared with predictions
from the (shape-function) OPE, after correcting for acceptances and detector effects.
Each plot in the figure shows the following three results for the indicated MC event sample: 
the detector-level signal distributions and the total number of 
events (TP+FP) accepted by the given BNN (upper panels), and the signal 
acceptance of the BNN (lower panels), all normalized to the number of detector-level signal events. 
The left (right) column uses \NNone  (\NNtwo). 
The BNNs are trained on \Evtgen data, and then tested on both \Evtgen and \Sherpa data.  
For \NNtwo, we also display the background acceptance in the lower panels,
using the scale for the $y$-axis displayed on the right of the plots. The background acceptance 
for \NNone is negligible across phase space and is thus not shown.

The figure highlights an inevitable fact --  since \NNone uses kinematic features to discriminate between
the signal and background, its acceptance is kinematics dependent.  The acceptance is higher
in the theoretically background-free regions of low $M_X$, high $q^2$, and high $p_\ell^*$, and lower 
in regions where the charm background is large.  

It is interesting and important to study 
the MC-data dependence of the signal acceptance in these two regions, and connect it to kinematic
modelling uncertainties in the MCs.  Take for example the results as a function of $M_X$  in the top left of the figure.  In the $0<M_X<1.5$~GeV bin, the \Evtgen and \Sherpa modelling of the $b\to u$ signal differ dramatically, with far more events in the \Sherpa sample, and also a very different shape as seen in the finely binned distributions shown in Figure~\ref{fig:sherpa_vs_evtgen_high_level_features}.  This is not entirely unreasonable, as the details of the low-$M_X$ distributions  depend on the method for matching resonant and non-resonant modes, and  even the integrated distribution over the entire bin depends on the exact implementation of the shape-function OPE.  However, the MC-dependence of the signal distribution in this theoretically intricate region does not propagate into the signal acceptance of 
\NNone, which is essentially MC-independent. 

Contrast this with the high-$M_X$ region, especially in the bins above 1.9~GeV where the charm
background is large. In this case, the marked difference in the shapes of the \Evtgen and \Sherpa signals 
as a function of $M_X$ does lead to noticeably different signal acceptances. On the other hand, kinematic distributions in the high-$M_X$ region where this becomes most significant are reliably calculable within the local OPE (before detector effects), so the MC-dependence can be viewed as an improvable deficiency in the current implementation of \Sherpa, which does not perform a matching with first-principle predictions  as described in Section~\ref{sec:EvtGenVsSherpa}, rather than as an irreducible kinematic modelling uncertainty. One would therefore expect a reasonable MC uncertainty associated with extrapolating the accepted events to the full fiducial region, although this deserves careful quantitative study in actual experimental analyses.

Similar qualitative comments hold for the $p_\ell^*$ and $q^2$ distributions -- the signal acceptances are essentially MC-independent in the highest bins, where kinematic modelling dependence due to non-perturbative shape-function effects is expected to be significant, but then start to become MC-dependent in the lower bins, where the local OPE is applicable.  On the other hand, the acceptances are somewhat flatter in these variables than in $M_X$, never dropping below 60\% in any of the bins.

The exclusion of kinematic input features from \NNtwo leads to a different qualitative 
picture of event acceptance compared to \NNone.  The right-hand side of 
Figure~\ref{fig:high_extra_acceptance} shows that its signal acceptance as a function
of $M_X$ is considerably flatter, remaining large at and above the $m_D$ resonance, 
although at the price of rejecting far less background. In total,  \NNtwo also accepts less of the signal. 
Whereas \NNone accepts 75\% (85\%) of the \Evtgen (\Sherpa) signal, the corresponding numbers for \NNtwo are 61\% (53\%) at the value of the threshold classifier which optimizes the significance improvement.  
For the $q^2$ and $p_\ell^*$ distributions the acceptances of \NNtwo are only moderately flatter than \NNone, if at all. The signal acceptances of \NNtwo are reasonably independent of the MC testing data across the kinematic phase space.  However,  unlike \NNone, noticeable differences can
be seen in the lowest $M_X$ and highest $q^2$ and $p_\ell^*$ bins, where 
shape-function effects and kinematic modelling are expected to be most important.
The background acceptance of \NNtwo is relatively flat at high $M_X$ and low $p_\ell^*$, but not at low $q^2$. Moreover, in the lowest $M_X$ bins as well as the high-$q^2$ region the background is largely excluded; these regions correlate with a large missing mass squared.

These observations show that MC-dependence of the acceptances of a given BNN is 
subtle -- avoiding sensitivity to kinematic modelling by excluding kinematic features 
is not always possible.  As a further illustration, consider a NN, 
 \NNbinned, taking as input the following features
\begin{equation}
\begin{split}
  Q_{B_\text{tag}} , \quad
 &  \text{ID}_i , \quad
  Q_i, \quad 
  [q^2]_{\rm binned}, \quad  
  [M_X]_{\rm binned} ,\quad
    [p_\ell^{*}]_{\rm binned} ,\quad
    N_\ell ,\quad
    N_{K^\pm} ,\quad
    N_{K^{0}} , \\
    N_\text{hadron} , \quad
    &M_\text{miss}^2 , \quad
    Q_\text{tot} , \quad
   N_{\pi^0_\text{slow}}, \quad
    N_{\pi^\pm_\text{slow}} , \quad
    M_{\text{miss},\, D^*}^2(\pi^0_\text{slow}) , \quad
    M_{\text{miss},\, D^*}^2(\pi^\pm_\text{slow}) .
\end{split}
\label{eq:features_binned}
\end{equation}
\NNbinned is the same as \NNone, except that particle 4-momenta  are
excluded\footnote{The high-level features for \NNbinned also differ from \NNone in that $P_+$ 
is included in the latter case but not the former. We verified that adding or taking it away
from makes a negligible numerical difference.}, and the high-level kinematic features 
are defined in the bins 
\begin{equation}
    \begin{split}
        M_X &= [0, \, 1.4, \, 1.6, \, 1.8, \, 2, \, 2.5,\, 3,\, 3.5] \, \gev \\
        p_\ell^* &=  [ 1, \, 1.25, \, 1.5, \, 1.75, \, 2, \, 2.25, \, 3] \, \gev \\
        q^2 &= [0, \, 2.5, \, 5, \, 7.5, \, 10,\, 12.5,\, 15,\, 20,\, 25] \, \gev^2 \, .
    \end{split}
\end{equation}
This binning matches that used in the construction of the hybrid Monte Carlo implemented within \Evtgen in Ref.~\cite{Cao:2021xqf}, and is sufficiently wide that fully inclusive distributions within these bins are accessible to the (shape-function) OPE.  In other words, unlike \NNone, this set-up is blind to the
heavily model-dependent point-by-point distributions of the hybrid Monte Carlo in the 
low $M_X$ and high $p_\ell$ and $q^2$ region, at least as far as the explicit input features are concerned. 

\begin{figure}
    \centering
    \includegraphics[width=0.32\textwidth]{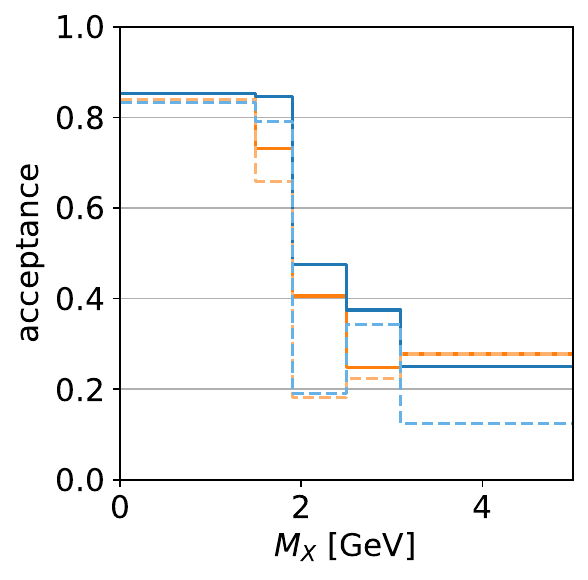}
    \includegraphics[width=0.32\textwidth]{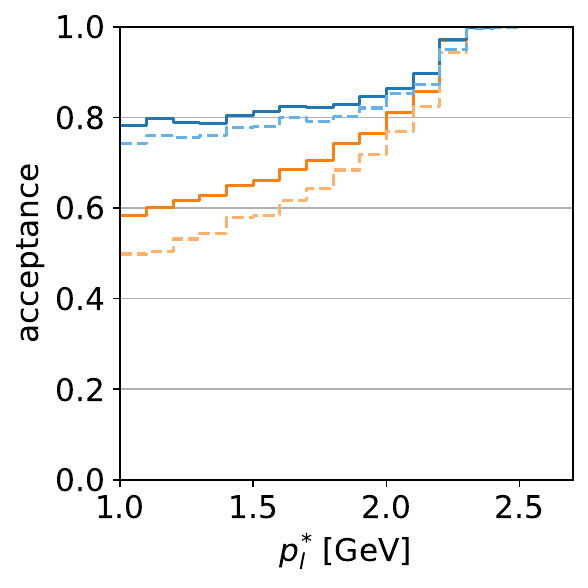}
    \includegraphics[width=0.32\textwidth]{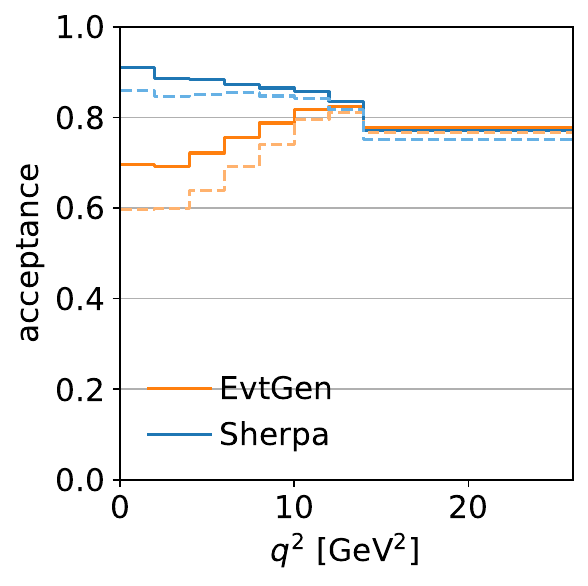}
    \caption{Signal acceptance as a function of $M_X$, $p_\ell^*$ and $q^2$ for \NNone (solid lines) compared to \NNbinned (dashed lines) defined in Eq.~\eqref{eq:features_high}. }
    \label{fig:acceptance_binned_vs_continuous}
\end{figure}

In Figure~\ref{fig:acceptance_binned_vs_continuous} we compare the acceptances
of \NNone and \NNbinned as a function of kinematic variables, using the same binning as in Figure~\ref{fig:high_extra_acceptance}.  Examining the figure shows that the MC-dependence of the \NNbinned acceptances are not reduced compared to \NNone, and 
they depend more strongly on the kinematic variables. In particular, when viewed as a function 
of $M_X$, \NNbinned shows a considerable drop in classification power in the higher bins,
where kinematic modelling uncertainties are expected to be best under control as long as the 
hybrid Monte Carlo is matched to OPE predictions.  Moreover, the maximal significance 
improvement $\hat{\sigma}$ drops: when tested on \Evtgen data \NNone has $\hat{\sigma}=5.67$ while 
\NNbinned has $\hat{\sigma}=5.46$.  It is thus far from clear that using a set-up such as 
\NNbinned would lead to a reduced theory uncertainty in $|V_{ub}|$ extractions compared
to \NNone, even though its explicit kinematic input features can be calculated within the 
(shape-function) OPE.

\subsection{Inclusivity in hadronic final states}
\label{sec:hadronization}

\begin{figure}[thb]
  \centering
    \includegraphics[height=5.5cm]{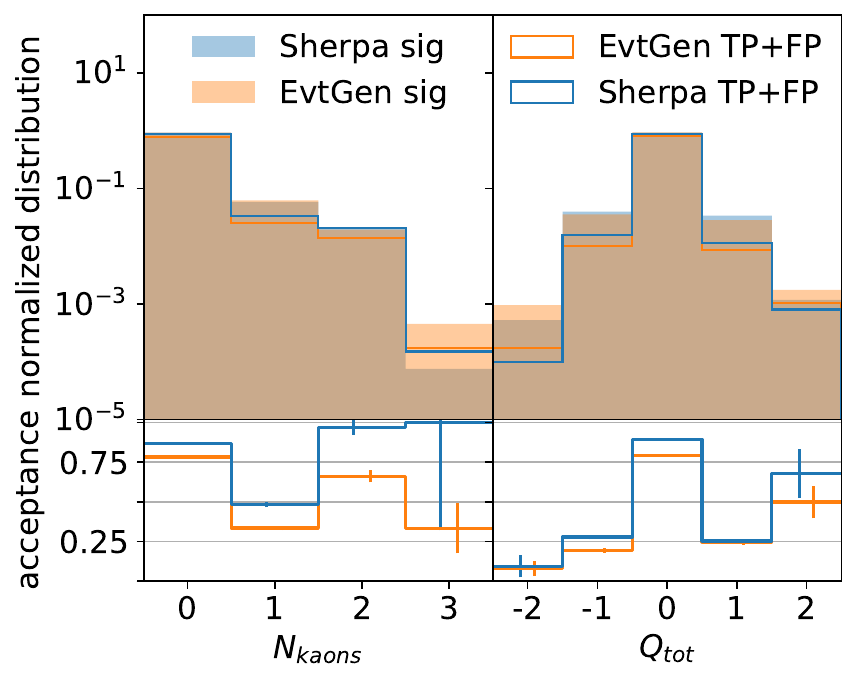}
    \;
    \includegraphics[height=5.5cm]{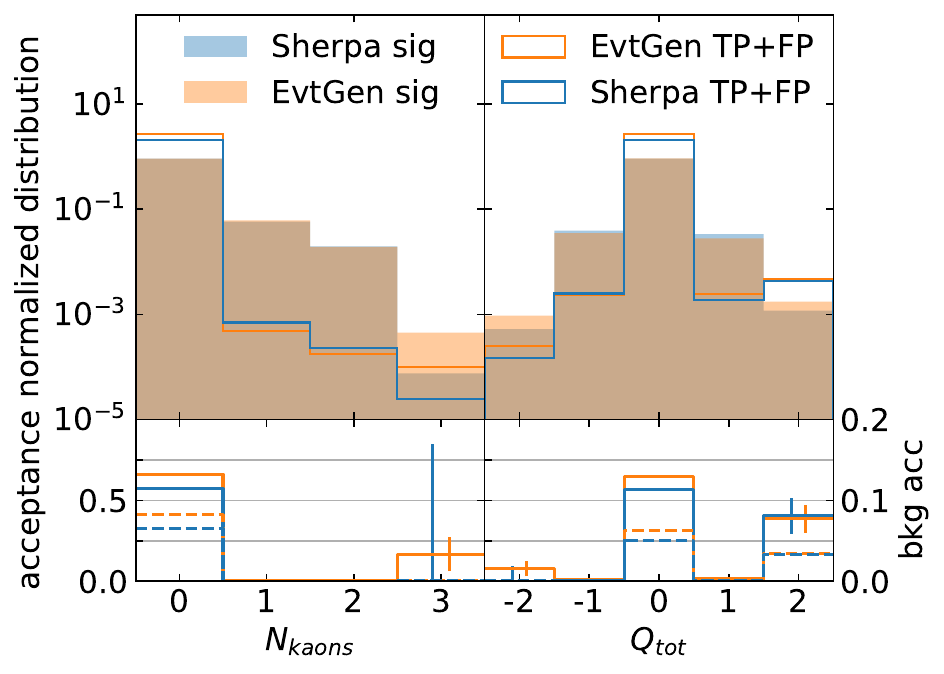}
  \caption{$Q_{\text{tot}}$ and $N_\text{kaons}$ distributions and signal acceptance for \NNone (left) and \NNtwo (right) trained on \Evtgen data. 
For \NNtwo the dashed lines in the lower panels show the background acceptance, using the scale for the $y$-axis displayed on the right.
In the lower panels, error bars highlight the MC uncertainty on the acceptance which for most bins (all bins for the background acceptance) is too small to be visible in the plots.}
  \label{fig:sherpa_trained_vs_evtgen_trained_global}
\end{figure}

We now shift our focus to inclusivity in properties of the final-state  $X_u$ system
which appear only after fragmentation into hadrons.  Such 
features are by definition inaccessible to OPE-based QCD calculations, which rely on a sum
over hadronic final states in order for quark-gluon duality to apply.  

In Figure~\ref{fig:sherpa_trained_vs_evtgen_trained_global} we display the same information 
as in Figure~\ref{fig:high_extra_acceptance}, but this time as a function of the number of kaons and total charge in the event. The number of kaons is an explicit probe of the flavour structure of the final state, whereas the total charge is closely related to the charged hadron multiplicity (see the discussion after
Eq.~(\ref{eq:features_high}) above).  Comparing the acceptance of \NNone and \NNtwo, we find that \NNtwo effectively vetos both signal and background events with kaons or a non-zero total 
charge.\footnote{The small contributions of events with $Q_\text{tot} =2$ to the total number of signal events is negligible.} 
 Therefore, when performing fits of the kinematic distributions after the \NNtwo analysis, a good understanding of both the signal and the charm background after strict cuts on the hadronic 
final states is required. 
\NNone, on the other hand, accepts a large proportion of events with kaons or a non-zero total charge and is thus more inclusive in (and  less dependent on) these hadronization-model dependent features.  

The number of signal events containing kaons in the final state is directly related to the $s\bar{s}$-popping probability~$\gamma_s$, which determines how often an $s\bar{s}$-pair is produced in the decay of the hadronic $X$~system.  It is interesting to further investigate the hadronization modelling sensitivity of the classifiers \NNone and \NNtwo resulting from their different kaon acceptances. Since the number of kaons in the background, which is entirely dominated by resonant contributions, is largely unaffected by changes of $\gamma_s$, we  investigate the sensitivity of the signal acceptance only.
We have produced additional \Evtgen test samples with a modified $s\bar{s}$-popping probability in the range $\gamma_s \in [0.1, \, 0.4]$ and apply \NNone and \NNtwo to these.\footnote{The tested $\gamma_s$ range is chosen to reflect the relatively large uncertainty on $\gamma_s$. The TASSO~\cite{Althoff:1984iz} and JADE~\cite{Bartel:1983qp} collaborations have experimentally determined the $s\bar{s}$-popping probability at center-of-mass energies of $12\,\gev$ and $27 \,\gev$ to be $\gamma_s = 0.35 \pm 0.05$ and $\gamma_s = 0.27 \pm 0.06$, respectively. 
The default \Pythia setting, resulting from a global tune of multiple fragmentation parameters, is 
$\gamma_s = 0.217$~\cite{Skands:2014pea}.}

\begin{figure}[t]
    \centering
    \includegraphics[width=0.45\textwidth]{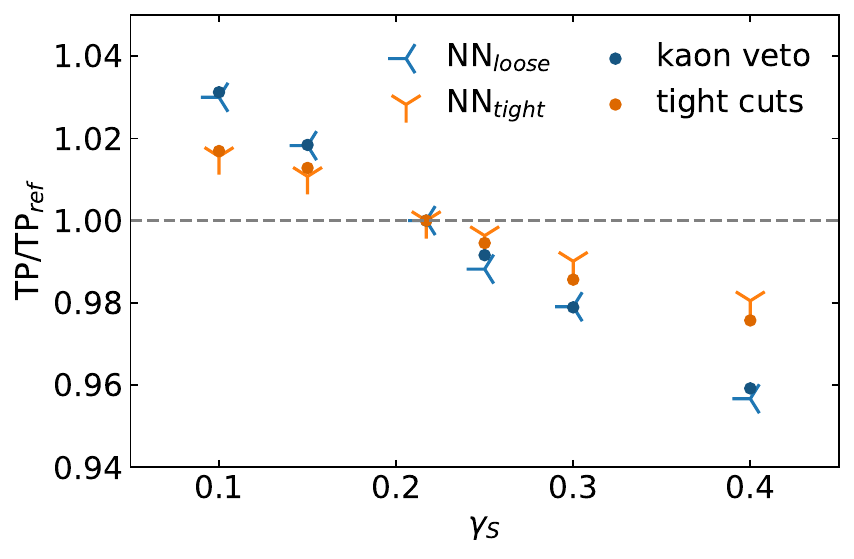} 
    \caption{Sensitivity of the number of TP events to the $s\Bar{s}$-popping probability $\gamma_s$. 
    The number of TP events at the \Pythia default is chosen as a reference value for each of the considered ML and cut-and-count approaches, $\TP_\text{ref} = \TP(\gamma_s = 0.217)$. The \textit{tight} cuts are defined by the cuts listed in Eq.~\eqref{eq:cuts_cut_and_count} plus $M_X<1.5 \,\gev$.}
    \label{fig:gammaS}
\end{figure}
In Figure~\ref{fig:gammaS}, we display the relative change of the number of TP events as a function of $\gamma_s$, taking the \Pythia default $\gamma_s = 0.217$~\cite{Skands:2014pea} as our reference value.  As events containing kaons are more likely to be classified as background by the BNNs, the number of TP events decreases with an increasing value of $\gamma_s$. For \NNtwo, which relies more heavily on the number of kaons as a features, the decrease of the signal acceptance is stronger. 

We contrast the effect of $\gamma_s$ on our ML analysis with a simple kaon veto as well as a cut-based approach defined by the cuts listed in Eq.~\eqref{eq:cuts_cut_and_count} plus an additional cut $M_X < 1.5 \, \gev$ (tight cuts).  The ML approach \NNtwo shows the same influence on $\gamma_s$ as a kaon veto, as expected from the signal acceptance shown in Figure~\ref{fig:sherpa_trained_vs_evtgen_trained_global}.
\NNone, however, is less disturbed by an increased value of $\gamma_s$ than its cut-and-count counterpart as it does not apply a stringent veto on kaons in signal events. 
Overall, our findings highlight the ability of ML approaches to lift the weight from single observables.

\subsection{Discussion}

The above results show that conclusions on the inclusivity of  \NNone and \NNtwo are based
heavily on how one thinks about the issue.  If the focus is on a flat coverage of kinematic
phase space, especially as a function of $M_X$, 
then \NNtwo, which does not include kinematic features, would be preferable.
If on the other hand one wishes to be more inclusive in the sum over exclusive hadronic 
final states on which quark-gluon duality is based, then \NNone, which accepts more 
events overall due to its increased discriminating power, is more attractive.

An important thing to keep in mind when considering \vub extractions is that in both cases MC modelling is used to extrapolate the signal from the fiducial region singled out by the 
NN to the partial inclusive branching fractions with a baseline kinematic cut of 
$p_\ell^{*}>1.0\,\gev$ (with no restrictions on the hadronic decomposition of the $X_u$ final state). 
For \NNone this extrapolation is mainly sensitive to the shape of the signal distribution at relatively
high $M_X$, which can reliably be calculated in the local OPE. For \NNtwo it is mainly
sensitive to non-perturbative phenomena such as the flavour decomposition and multiplicity 
of the hadronic final state across all kinematics. Given that the extrapolations are 
sensitive to different effects, it may be wise to pursue both approaches in real-life $|V_{ub}|$ extractions.

It is worth mentioning that the signal acceptance of the kinematics independent ``background suppression'' BDT used in the recent analysis of Ref.~\cite{Cao:2021xqf} is significantly smaller than that found using \NNtwo and our in-house detector simulation, so that the extrapolation from the accepted fiducial region to fully inclusive partial branching fractions with kinematic cuts is correspondingly larger. By the same token, we expect that the acceptance of \NNone in the high-$M_X$ region would be considerably lower in the full experimental environment, again requiring a larger extrapolation than seen
in our simplified set-up.

\FloatBarrier
\section{Conclusions}
\label{sec:conclusion}

We have performed a systematic study on the use of ML techniques in inclusive \vub determinations. While our analysis is based on a simplified set-up using an in-house detector 
simulation and seeking only to separate the $B\to X_u\ell\nu$ signal from the $B\to X_c\ell\nu$
background, it has revealed several important qualitative points.

First, in Section~\ref{sec:BDTvsNN}, we showed that using a deep neural network trained on low-level single-particle features leads to a small performance increase with respect to a BDT analysis based on high-level features of the type used in the Belle analysis~\cite{Urquijo:2009tp}.  While upgrading such analyses to modern ML standards
is certainly worthwhile, the modest performance increase produced by the more sophisticated
ML architecture implies that  the high-level features used in current 
BDTs are well-chosen --  the most important 
aspects of discriminating the $b\to u$ signal from the $b\to c$ background can be understood
with physicist-engineered observables.

Second, in Section~\ref{sec:inclusivity} we studied the inclusivity of the fiducial region selected
by cuts on the classifier output of two types of neural networks: \NNone, based on input features of both kinematic and hadron-level features of the final states, such as the one just described and used in Ref.~\cite{Urquijo:2009tp}, and \NNtwo, which excludes the kinematic properties and is similar
to the BDT used in the recent analysis in Ref.~\cite{Cao:2021xqf}.  While the signal acceptance of 
\NNtwo is fairly flat across the kinematic phase space, it effectively makes hard cuts
in hadronic properties of the event such as the number of kaons and the total charge.
On the other hand, \NNone is significantly more inclusive in the hadronic decomposition of the final state and also in general, but  tends to give less weight to kinematic regions where there is a large overlap with the $b\to c$ background.  Both of these issues deserve careful consideration when
assessing systematic theory uncertainties related to MC extrapolation from the  
fiducial regions to partial branching fractions that are calculable within the (shape-function) 
OPE in QCD.  

Finally, as the Belle~II measurements become systematics dominated, it will be important 
to pay close attention to the sensitivity of supervised ML approaches to the MC data on which
they are trained. We have investigated the influence of a modified $s\bar{s}$-popping probability on the signal acceptance using \Evtgen data. A ML approach based on kinematic information, such as \NNone, is generally less biased by changes of global event parameters. Furthermore, in 
Section~\ref{sec:setup} we showed results from the multipurpose MC event
generator \Sherpa in addition to those from  \Evtgen, which has been the exclusive MC tool
for all previous  \vub analyses, and in Section~\ref{sec:inclusivity} we discussed features
appearing when the BNNs were trained and tested on event sets produced by different MCs.
While \Sherpa needs optimisation in matching with OPE-based theory predictions before it
can be used in experimental analyses, investigating the stability of ML approaches against MCs whose modelling is based on different theory assumptions can provide a powerful stress-test
on MC uncertainties, beyond the current practice of exploring modifications within \Evtgen.

\section*{Acknowledgements}
We thank Florian Bernlochner, Tim Gershon, Frank Krauss, Michel Luchmann and Marcello Rotondo for useful discussions. 
A.B.~gratefully acknowledges support from the Alexander-von-Humboldt foundation as a Feodor Lynen Fellow. KW.K.~is supported by the UK Science and Technology Facilities Council (STFC) under grant ST/P001246/1. B.P.~is grateful to the Weizmann Institute of Science for its kind hospitality and support through the SRITP and the Benoziyo Endowment Fund for the Advancement of Science.

\appendix
\newpage

\section{Detector simulation}
\label{app:detector}

Theoretically, the signal and background processes are well separated by the through kinematic boundaries at $M_X = m_D$, $P_+ = m_D^2/m_B$ and $p_\ell^* =(m_B^2-m_D^2)/(2 m_B)$.
However, detector effects lead to large contributions from the $B \to X_c \ell \nu$ background to the $B \to X_u \ell \nu$ signal region, and it is necessary to include them in order to mimic the challenges of the experimental environment. 

In the following, we describe our in-house detector simulation meant to capture the main
features of a more complete one. We list the assumed parameters 
for detector resolution in Section~\ref{subsec:resolution}
and for detector efficiencies and mistagging probabilities 
in Section~\ref{subsec:efficiencies}.
Most of these values are based on the description of the 
BaBar detector in Ref.~\cite{TheBABAR:2013jta}, 
from the BaBar analysis of the inclusive determination of $\vub$ paper~\cite{Lees:2011fv} and the corresponding PhD thesis on the same subject~\cite{thesis_Gagliardi}.
We compare the resulting distributions after our detector simulation to those shown in the recent reanalysis of Belle events in Ref.~\cite{Cao:2021xqf}. 
We highlight that the beam energies in Belle ($3.5\,\gev$ and $8.0\,\gev$) are slightly different from the values we used in our MC event generation ($4.0\,\gev$ and $7.0\,\gev$), see Section~\ref{sec:setup}. We therefore expect deviations of the lab-frame momenta on the level of $\lesssim 10\,\%$.

\subsection{Detector resolution}
\label{subsec:resolution}

We assume perfect reconstruction of the direction of each detected particle and we only smear the energy (momemtum) for photons (charged particles). 

The energy resolution of photons is parametrized by~\cite{thesis_Gagliardi}
\begin{align}
	\frac{\sigma_{E_\gamma}}{E_\gamma} = \frac{2.32 \,\%}{E_\gamma^{1/4}} \oplus 1.85 \,\%, \quad E_\gamma \text{ in GeV.}
\end{align}

For the resolution of charged particles, we use the $p_T$ resolution of the Drift Chamber (DCH) 
which is the main tracking device for charged particles with $p_T \geq 120\,\mev$~\cite{thesis_Gagliardi}.
\begin{align}
	\frac{\sigma_{p_T}}{p_T} = 0.45\,\% \oplus 0.13\, \% \, p_T, \quad p_T \text{ in GeV.}
\end{align}
We apply this formula on all charged particles, also those with $p_T < 120\,\mev$.

\subsection{Efficiencies and mistagging}
\label{subsec:efficiencies}
For charged particles/tracks, the overall reconstruction efficiency is 
$98\,\%$ for momenta $p \geq 200\,\mev$ (DCH)~\cite{thesis_Gagliardi}.

We assume that mistagging is only relevant for 
\begin{equation*}
\begin{aligned}[t]
 \text{true } \pi^{\pm} &\rightarrow \text{ fake } K^{\pm}\\
                                  &\rightarrow  \text{ fake } e \\
                                  &\rightarrow  \text{ fake } \mu
\end{aligned}
\qquad \qquad
\begin{aligned}[t]
 \text{true } K^{\pm} &\rightarrow \text{ fake } \pi^{\pm}\\
                                 &\rightarrow  \text{ fake } e
\end{aligned}
\end{equation*}

\subsubsection*{Photons}

Photons are detected with an efficiency of $96 \,\%$ for energies above $20\,$MeV.
\begin{align}
	\text{eff}_\gamma (E_\gamma) =  0.96 \, (E_\gamma \geq 0.02), \quad E_\gamma \text{ in GeV}
\end{align}

\subsubsection*{Electrons}
Electrons need to have a minimum momentum of $p_\text{lab} = 500 \,\mev$ in the lab frame. 
Their efficiency is $93\,\%$ above this threshold~\cite{Lees:2011fv}.
\begin{align}
	\text{eff}_e(p) =  0.93 \, (p \geq 0.5), \quad p \text{ in GeV}
\end{align}

\subsubsection*{Muons}

Muons need to have a minimum momentum of $p_\text{lab} = 500\,\mev$ in the lab frame. 
Their efficiency is $90\,\%$ above this threshold.
\begin{align}
	\text{eff}_e(p) = 0.9 \,  (p \geq 0.5), \quad p \text{ in GeV}
\end{align}
Since muons and electrons/hadrons are detected in different detector parts, we assume the muon fake rate for electrons and hadrons to be negligible. 

\subsubsection*{Kaons}
Charged kaons need to have minimum momenta of  $p_\text{lab} \geq 300\,\mev$ to be identified. 
The efficiency is taken from Figure~3.5 of Ref.~\cite{thesis_Gagliardi}. It drops linearly 
for momenta satisfying $p<7\,\gev$, at values above this we approximate the efficiency using a quadratic function:
\begin{align}
\text{eff}_{K^{\pm}}(p) =
\begin{cases}
        0,      & p < 0.3\\
       -0.8\, p+1.23 ,  & 0.3 \leq p  < 0.7  \qquad   p \text{ in GeV}\\
        0.86-0.35(p-1.5)^2 , & 0.7 \leq  p < 1.8 \\
       -0.0225 \, p+0.87 , & p > 1.8
\end{cases}
\end{align}     

We determine possible $K_s^0$ candidates based on the invariant mass of opposite-sign pion pairs.
Pairs in the mass range $m_{\pi^+\pi^-} \in [0.490, \, 0.505]\,\gev$ are assumed to result from $K_s^0$ decays with a $40\,\%$ probability,
see Figure~3.6 of Ref.~\cite{thesis_Gagliardi}.

We model the misidentification of kaons as electron as
\begin{align}
\text{mis}_{e | K}(p) =
\Bigg\{ 
\begin{array}{lcc}
        0,      &  p < 0.05\\
      0.004-0.001 \, p, & 0.05 \leq p  < 4.0 & \quad  p \text{ in GeV}\\
        0 , & p > 4.0
        \end{array}
\end{align}     

\subsubsection*{Pions}
For the reconstruction efficiency of slow, i.e.\ low momentum, pions we use the values given 
in Ref.~\cite{Aubert:2000bz}. The efficiency for pions grows exponentially from  $20\,\%$ at $p_T=50\,\mev$ to
$80\,\%$ at $p_T=70\,\mev$, see also Figure~9 of Ref.~\cite{Aubert:2000bz}.
For pion momenta $p\geq 0.4\,\gev$, we assume the reconstruction efficiency to 
drop linearly, compare Figure~89 of Ref.~\cite{TheBABAR:2013jta}.
\begin{align}
\text{eff}_\pi(p) =
\Bigg\{ 
\begin{array}{lcc}
        0,      &  p < 0.05\\
       1 -13 \exp(-86.29\, p+560.4\, p^2-1601\, p^3+1625 \, p^4), & 0.05 \leq p  < 0.4 & \quad  p \text{ in GeV}\\
        1-0.015 \, p, & p > 0.4
        \end{array}
\end{align}     

The efficiency for pions to be misidentified as kaons is taken from Figure~3.5 of Ref.~\cite{thesis_Gagliardi}. We 
approximate the momentum dependence as linear for low momenta and constant for larger momenta
\begin{align}
\text{mis}_{K | \pi}(p) =
\Bigg\{ 
\begin{array}{lcc}
        0,      &  p < 0.05\\
       0.01 \, p , & 0.05 \leq p  < 2.0 & \quad  p \text{ in GeV}\\
        0.02 , & p > 2.0
        \end{array}
\end{align}     

We assume the efficiency for pions to be misidentified as muons to be $0.5\,\%$ below $1\,\gev$ 
and $1\,\%$ above this value (Figure~3.4 in Ref.~\cite{thesis_Gagliardi}).
We do not model any angular dependence. 
\begin{align}
\text{mis}_{\mu | \pi}(p) =
\Bigg\{ 
\begin{array}{lcc}
        0,      &  p < 0.5\\
       0.005 \, p , & 0.5 \leq p  < 1.0 & \quad  p \text{ in GeV}\\
        0.1 , & p > 1.0
        \end{array}
\end{align}     

We model the misidentification of pions as electron as
\begin{align}
\text{mis}_{e | \pi}(p) =  0.001p \, ( p>0.5),  \quad  p \text{ in GeV}
\end{align}     

\subsection{Validation}
\begin{figure}[tbh]
\centering
	\includegraphics[width=.41\textwidth]{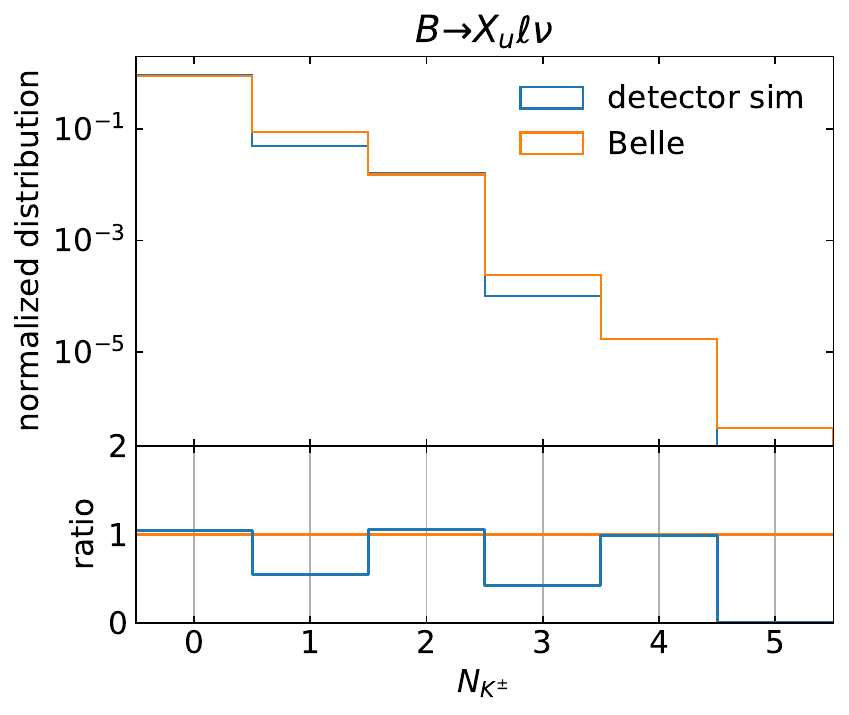}\quad
	\includegraphics[width=.41\textwidth]{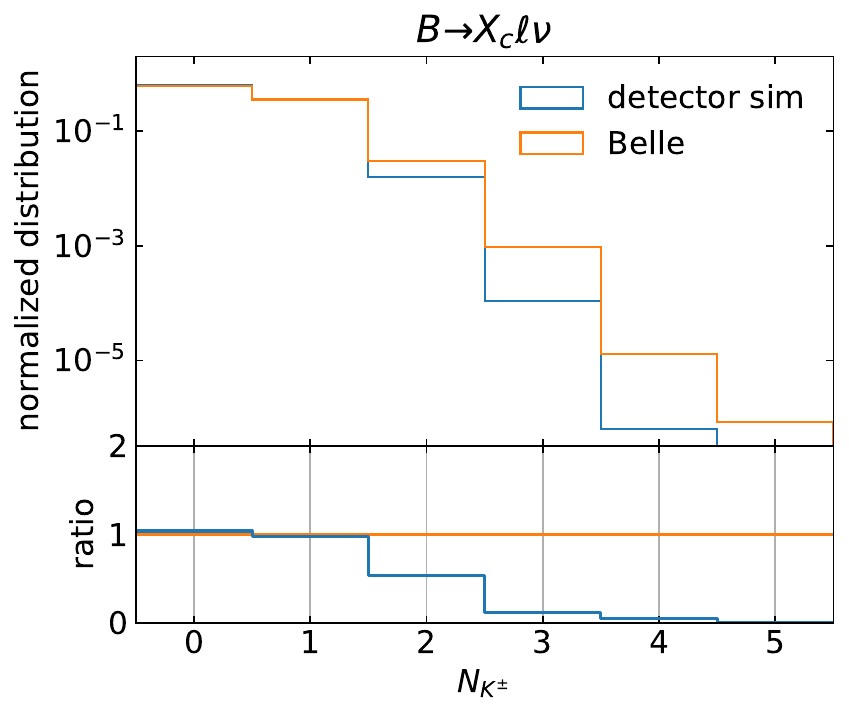}\\[2mm]
	\includegraphics[width=.41\textwidth]{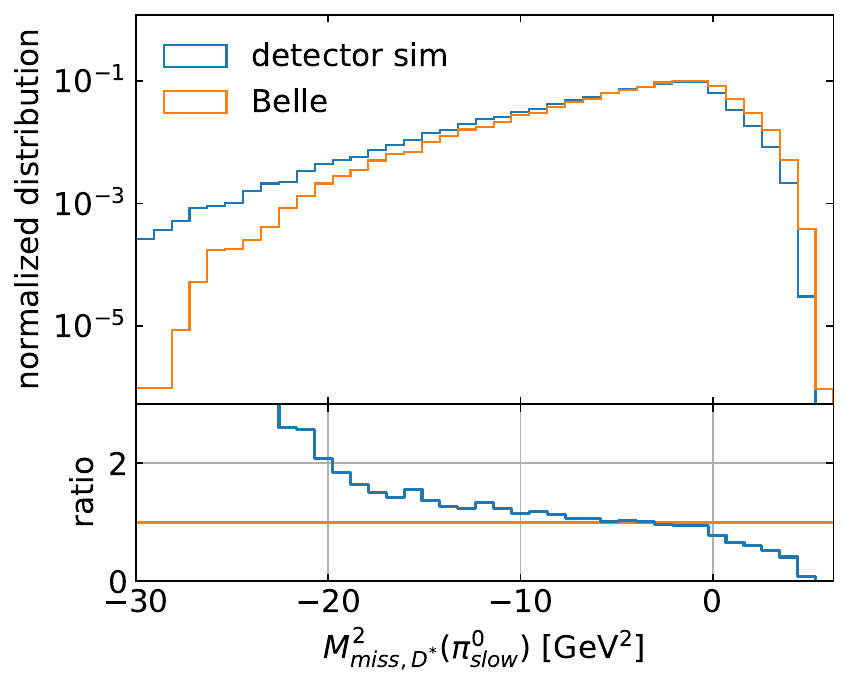}\quad
	\includegraphics[width=.41\textwidth]{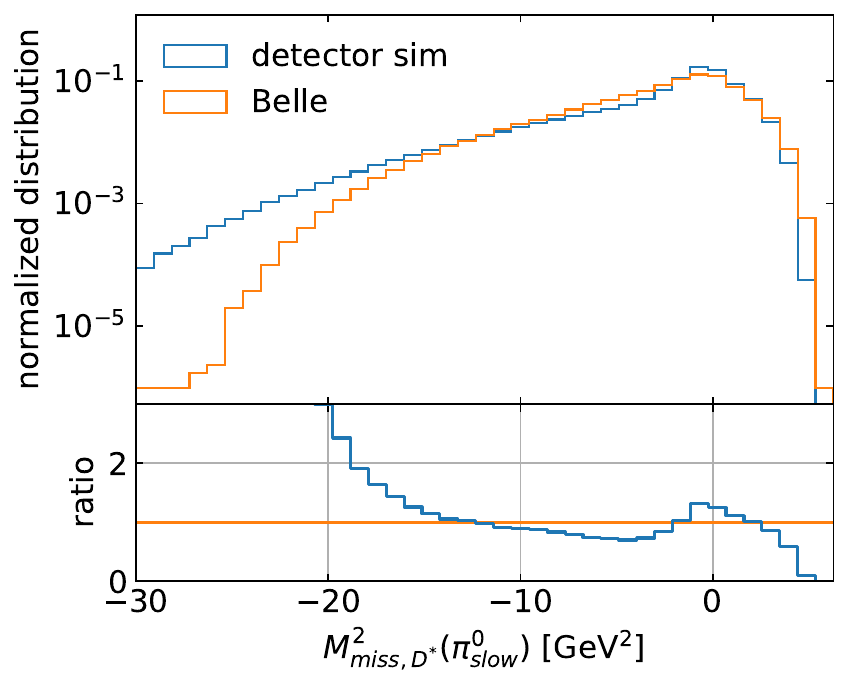}\\[2mm]
	\includegraphics[width=.41\textwidth]{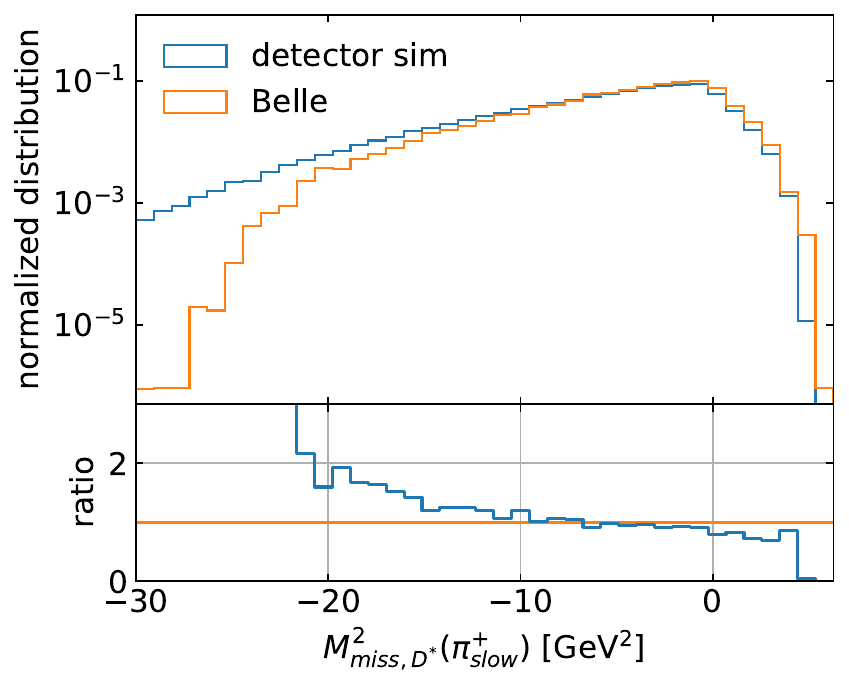}\quad
	\includegraphics[width=.41\textwidth]{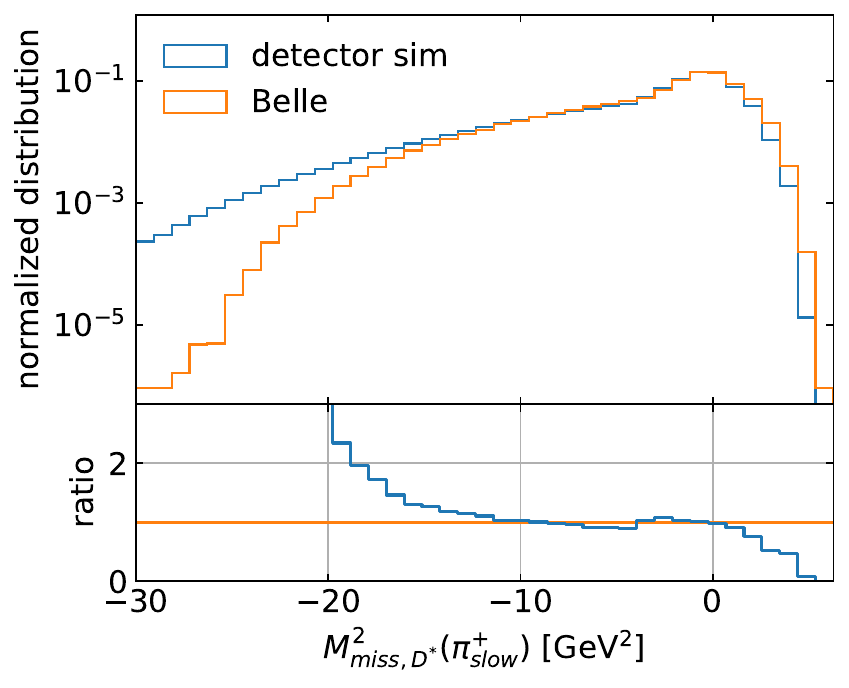}\\[2mm]
	\includegraphics[width=.41\textwidth]{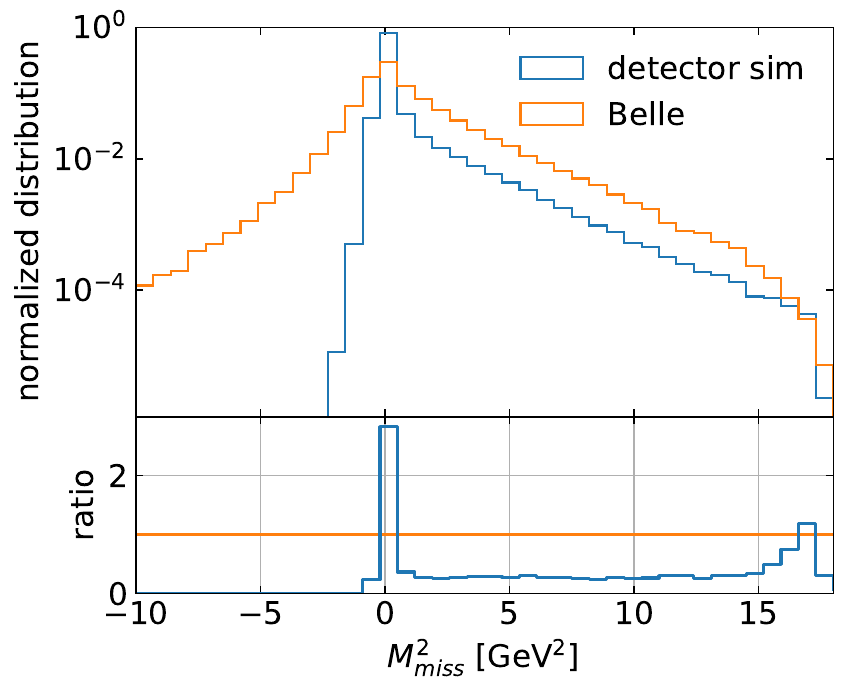}\quad
	\includegraphics[width=.41\textwidth]{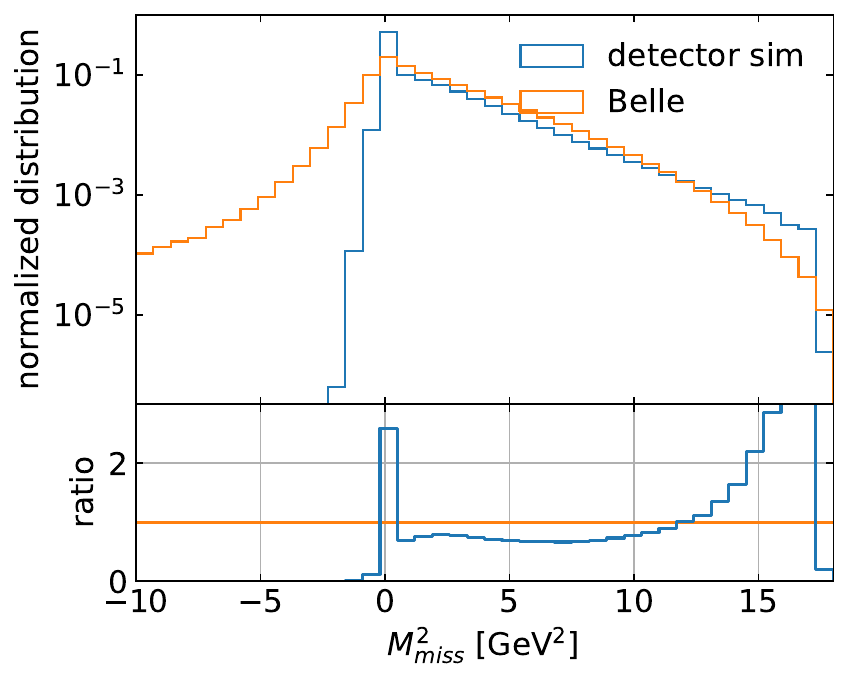}
	\caption{Detector simulation validation plots for signal (left) and background (right) contributions. 
	We compare the distributions of our 
	MC events after detector simulation (detector sim) with the MC events produced by the Belle collaboration displayed in Figure~14
	of Ref.~\protect\cite{Cao:2021xqf}.
	See paragraph below Eq.~\eqref{eq:features_high} for the feature definitions.}
	\label{fig:validation1}
\end{figure}

To validate our detector simulation, we reproduce Figure~14 of Ref.~\cite{Cao:2021xqf} in our Figure~\ref{fig:validation1}.
We find reasonable agreement for the number of charged kaons and the bulk of the $M_{\text{miss},\, D^*}^2(\pi_\text{slow})$ distributions.
Larger deviations between our detector simulation and the Belle values, for instance at low $M_{\text{miss},\, D^*}^2(\pi_\text{slow})$ or with a large number of kaons, appear in statistically much less relevant regions
and less than $2\,\%$ ($1\,\%$) of all signal (background) events lie at $M_{\text{miss},\, D^*}^2(\pi_\text{slow})< -20 \,\gev^2$. Less than $3\,\%$ of the background event contain more than one charged kaon.
Since we not include the effect of particles of the tagging side of the event being assigned to the signal side, we poorly underestimate the negative regime of the missing mass squared. 

\FloatBarrier

\subsection{Broader resolution}
\label{app:increased_resolution}
\begin{figure}[tbh]
\centering
    \includegraphics[width=.43\textwidth]{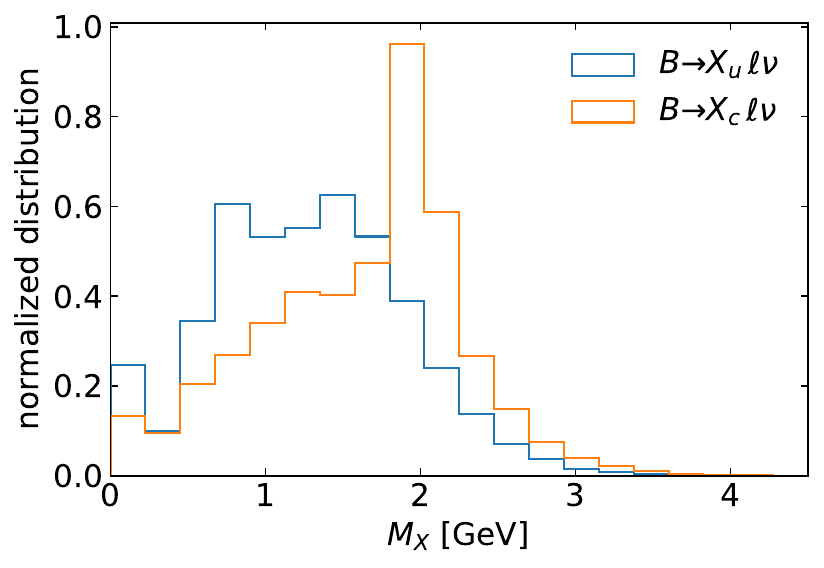}\quad
    \includegraphics[width=.43\textwidth]{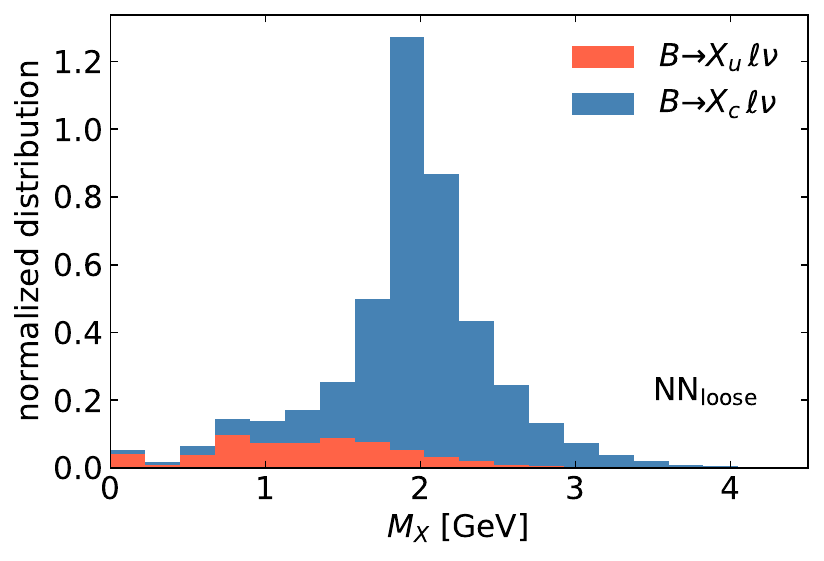}\\
	\includegraphics[width=.43\textwidth]{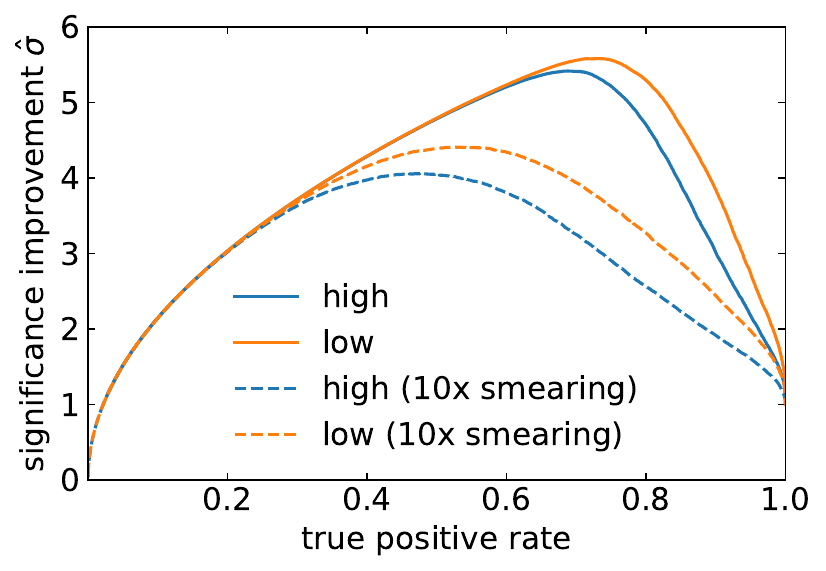}\quad
	\includegraphics[width=.43\textwidth]{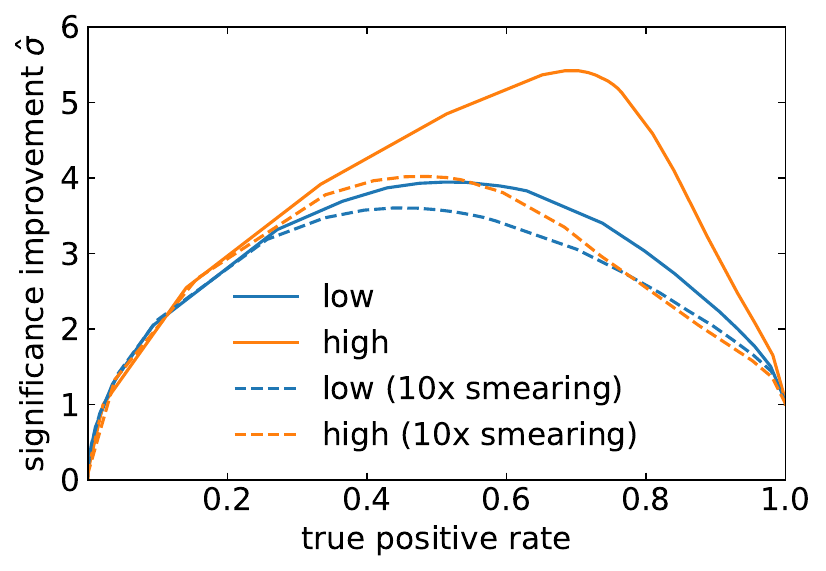}
	\caption{Top: $M_X$ distribution before (left) and after (right) the background suppression BDT (plot analogous to Fig.~6 in Ref.~[7]).
	Bottom: Significance improvement for a NN (left) and BDT (right) trained on high-level or low-level input features. We compare our standard set-up to training and testing on a sample with the detector resolution for photons and charged particles broadened by a factor 10.}
	\label{fig:smearing10}
\end{figure}

Some of the input features in our analysis do not fully resemble the experimental input features.  In this appendix, we study the dependence of our findings in Section~\ref{sec:Results} on the detector simulation. As a test case, we broaden the smearing of the charged particle momenta and photon energies. Increasing the smearing  by a factor $10$ brings the $M_X$ resolution to a level close to what is seen in experiment. The resulting $M_X$ distribution is shown in the top panel of Fig.~\ref{fig:smearing10}. 
The modified particle resolution will similarly affect low-level and high-level input features and allows us to study its impact on the different multivariate analysis set-ups. 
We re-perform our tight NN and BDT analyses using training and test data with the increased smearing and show the corresponding significance improvement of these analyses in the bottom panel of Fig.~\ref{fig:smearing10}. 
Qualitatively, the comparison of the high-level and low-level data sets has not changed. There are, however, some quantitative changes of the maximum significance reached.
For the NN the ratio of the maximum significance
$\hat{\sigma}(\text{NN}_\text{low})/\hat{\sigma}(\text{NN}_\text{high})$ 
changes from $1.03$ in the standard set-up to $1.09$ when increasing the smearing by a factor ten. 
For the BDT the ratio of the maximum significance $\hat{\sigma}(\text{BDT}_\text{high})/\hat{\sigma}(\text{BDT}_\text{low})$ 
changes from $1.38$ in the standard set-up to $1.12$ when increasing the smearing by a factor ten. 
In both cases the classifiers using high-level input features are more 
strongly affected than those using low-level features.
We would like to emphasise, however, that although broadening the detector resolution by a factor of ten brings the invariant mass resolution in line with that seen in experimental simulations, it is not a realistic scenario and therefore these results should be taken with a grain of salt.
\enlargethispage{10mm}

\section{Machine Learning analysis set-up}
\label{app:MVA_setup}

\subsection{Training and test sets}
\label{app:MVA_setup_test_train}
To train our classifiers, we create balanced data sets with $10$M $B\to X_u \ell \nu$ signal events and $10$M  $B\to X_c \ell \nu$ background events.
The data preparation process includes the application of the in-house detector simulation 
 and  a standard scaling of the data based on the training set.  Categorical features are one-hot encoded and are not scaled.
The training set is shuffled and $20\,\%$ of it is used for cross validation. 
For testing, we create two test sets with a physical signal-to-background ratio ($1/45$). 
Each test set contains $40$K signal and $1.8$M background events after detector simulation, which roughly corresponds to the number of semileptonic $B$-decays in a sample of 22.6M $B\bar{B}$ events.  

\subsection{Bayesian neural network}
\begin{table}[thb]
    \centering
    \begin{tabular}{lc}
        \toprule
         Bayesian neural network (BNN)  \\
         \midrule
         Input layer & number of features nodes\\
         1$^\text{st}$ hidden DenseFlipout layer & $256$ nodes, Sigmoid activation\\
         & batch normalisation\\
        2$^\text{nd}$ hidden DenseFlipout layer & $256$ nodes, Sigmoid activation\\
        & batch normalisation\\
        3$^\text{rd}$ hidden DenseFlipout layer & $256$ nodes, Sigmoid activation \\
        Output layer &1 node, Sigmoid activation \\
        Kernel posterior function & mean field normal distribution\\
        Bias posterior function & mean field normal distribution\\
        Kernel divergence function & KL divergence function\\
        \midrule
        Loss function & binary cross-entropy \\
        Optimizer & Adam \\
        learning rate & $0.1$ for first 10 epochs \\
        & then decreasing with $e^{-0.1}$ each epoch \\
        \bottomrule
    \end{tabular}
    \caption{Neural network architecture.}
    \label{tab:nn_setup}
\end{table}
Our Bayesian NN is implemented with \texttt{Tensorflow}~\cite{tensorflow2015-whitepaper}, \texttt{TensorFlow-Probability}~\cite{DBLP:journals/corr/abs-1711-10604} and \texttt{Keras}~\cite{chollet2015keras} with a total of 5 layers.
The number of nodes of the input layer is the number of input features. 
There are 3 hidden \texttt{DenseFlipout} layers~\cite{wen2018flipout}, each of them containing 256 nodes using the Kullback-Leibler (KL) divergence function as the kernel 
divergence function.  The KL divergence function is defined as
\begin{equation}
    \text{KL}[q(\omega),p(\omega|C)] = \int d\omega~q(\omega) \log \frac{q(\omega)}{p(\omega|C)}\, ,
    \label{eq:kl_div}
\end{equation}
where $p(\omega|C)$ is the posterior probability distribution given classifier $C$ and $q(\omega)$ is the approximation created through the classifier~\cite{alexgraves2011}. 
We use a sigmoid activation function for all hidden layers. The first two hidden layers are followed by a batch normalisation layer which scales the weights and biases to have mean $= 0$ and standard deviation $= 1$. This helps avoid the vanishing gradient problem with sigmoid functions.   
The output layer only has 1 node with a sigmoid activation function, the posterior function for the kernel and bias are both assumed to be mean field normal distributions. The kernel divergence function for the output layer is also the KL divergence function. 

We use binary cross-entropy as our loss function and apply the Adam~\cite{Kingma2015AdamAM} optimizer. The KL divergence is automatically added to the loss during training. Early stopping and model checkpoints are in place to monitor the validation loss of each epoch. The model weights from the best performing epoch are saved out and loaded back in before inference. We summarise the BNN architecture in Table~\ref{tab:nn_setup}.

\subsection{Boosted decision tree}
\label{sec:BDT}
\begin{table}[thb]
	\centering
\begin{tabular}{ lc} 
		\toprule
		 Boosted decision tree (BDT)  \\ 
		\midrule
		Classifier & XGBoost \\
		Max depth & $10$ \\
		 Learning rate & $0.4$ \\ 
		 Number of estimators & $300$\\ 
		 Gamma & 1\\
		Subsample & $0.9$\\
		Colsample$\_$bytree & $0.7$\\
		Loss function & logloss \\
		\bottomrule
\end{tabular}
\caption{Boosted decision tree architecture.}\label{tab:bdt_setup}
\end{table}
The BDT is implemented with \texttt{XGBoost}~\cite{Chen:2016:XST:2939672.2939785}. 
We allow for a maximum depth of $10$ as higher depth did not improve performance. The learning rate is fixed at $0.4$. The number of estimators is set to $300$ with early stopping in place. The gamma factor is fixed at~$1$. The subsample ratio of the training instance is $0.9$ and subsample ratio of columns when constructing each tree is set to be $0.7$ to reduce the risk of overfitting. The BDT set-up is summarized in Table.~\ref{tab:bdt_setup}.\\

In training the algorithms, the hyperparameters displayed in Tab.~\ref{tab:nn_setup} and~\ref{tab:bdt_setup} were predetermined with minimal optimization through \texttt{HyperOpt}~\cite{Bergstra_hyperopt:a}. 
\FloatBarrier
\newpage
\section{Plots of the high-level input features}
\label{app:morePlots}

\begin{figure}[thb]
  \centering
      \includegraphics[width=0.3\textwidth]{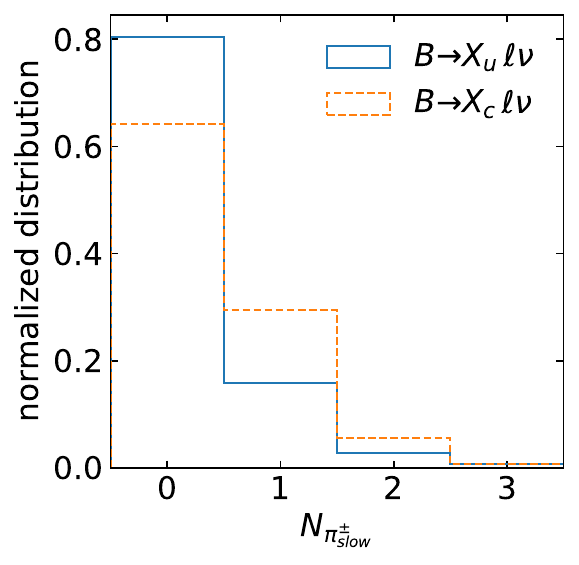}
    \;
    \includegraphics[width=0.3\textwidth]{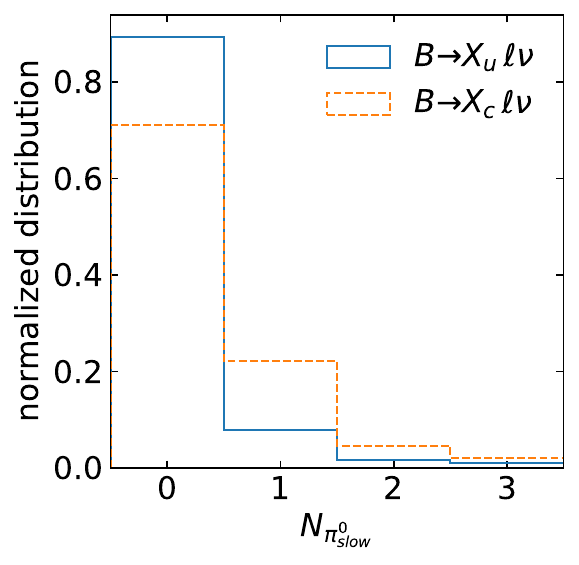}
    \;
    \includegraphics[width=0.3\textwidth]{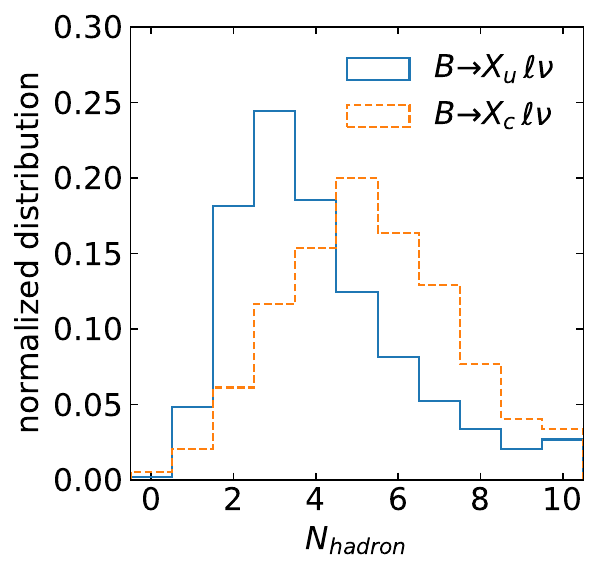}
    \\
    \includegraphics[width=0.3\textwidth]{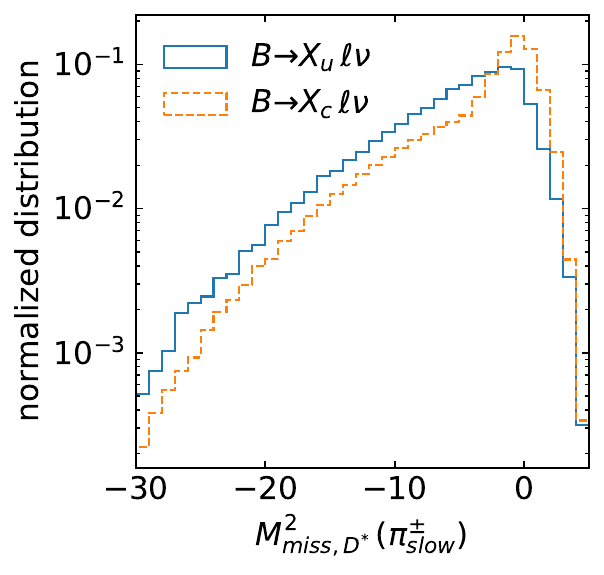}
    \;
    \includegraphics[width=0.3\textwidth]{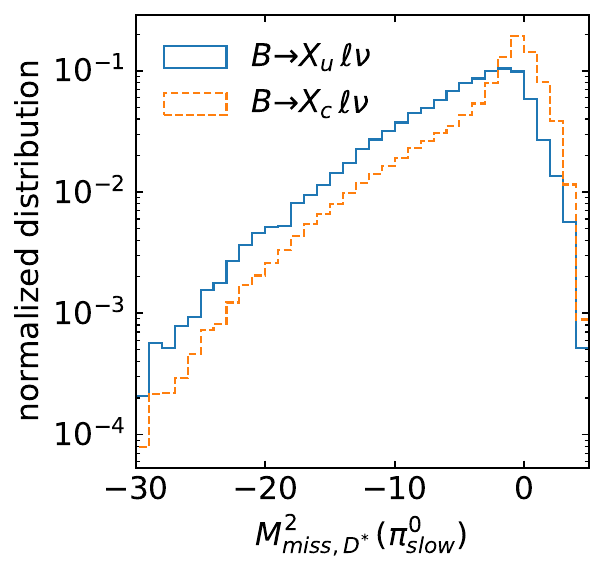}
    \; 
    \includegraphics[width=0.3\textwidth]{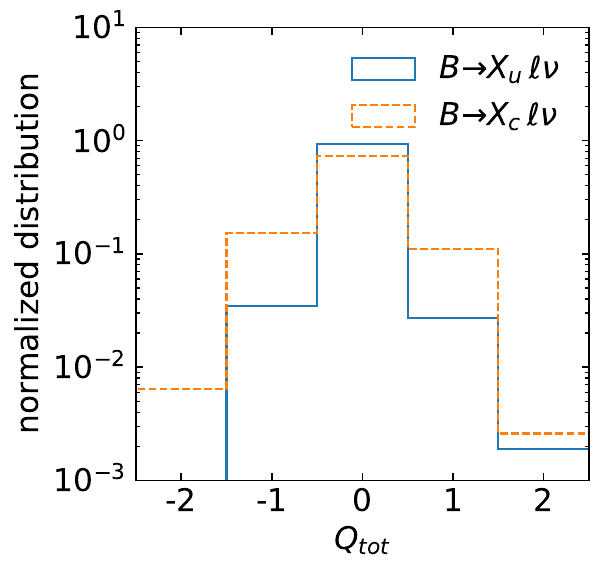}
    \\
    \includegraphics[width=0.31\textwidth]{figures/mm2_evt_afterDetector_aspect_log.pdf}
    
  \caption{Comparison of high-level features for $B\to X_u \ell \nu$ signal and $B\to X_c \ell \nu$ background events.}
  \label{fig:more_high_level_features}
\end{figure}
\FloatBarrier

\section{Training with \Sherpa}
\label{sec:Sherpa_Training}

\begin{figure}[thb]
    \includegraphics[height=5.5cm]{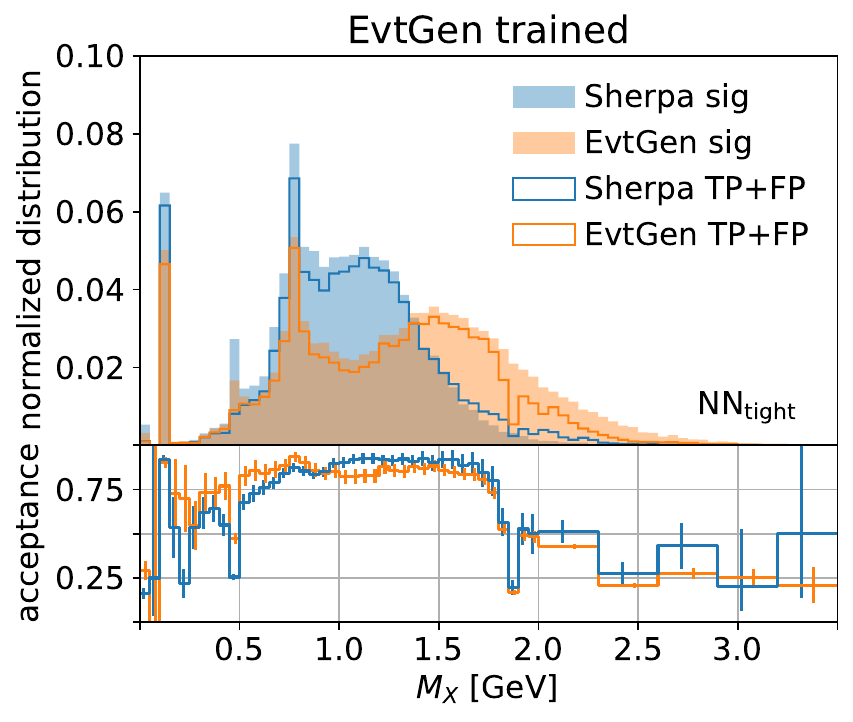}
    \qquad \;
     \includegraphics[height=5.5cm]{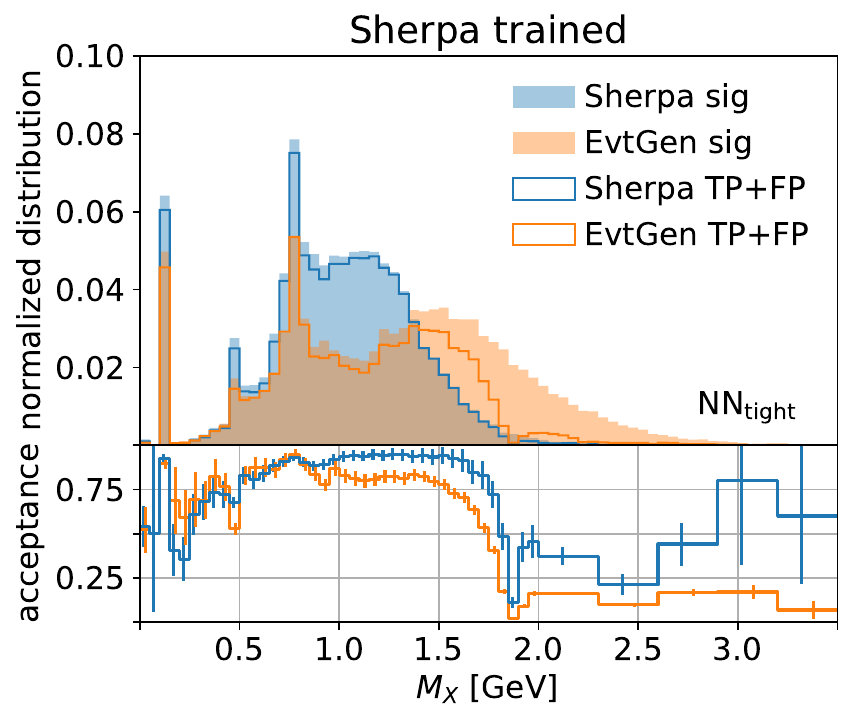}\\
        \includegraphics[height=5.2cm]{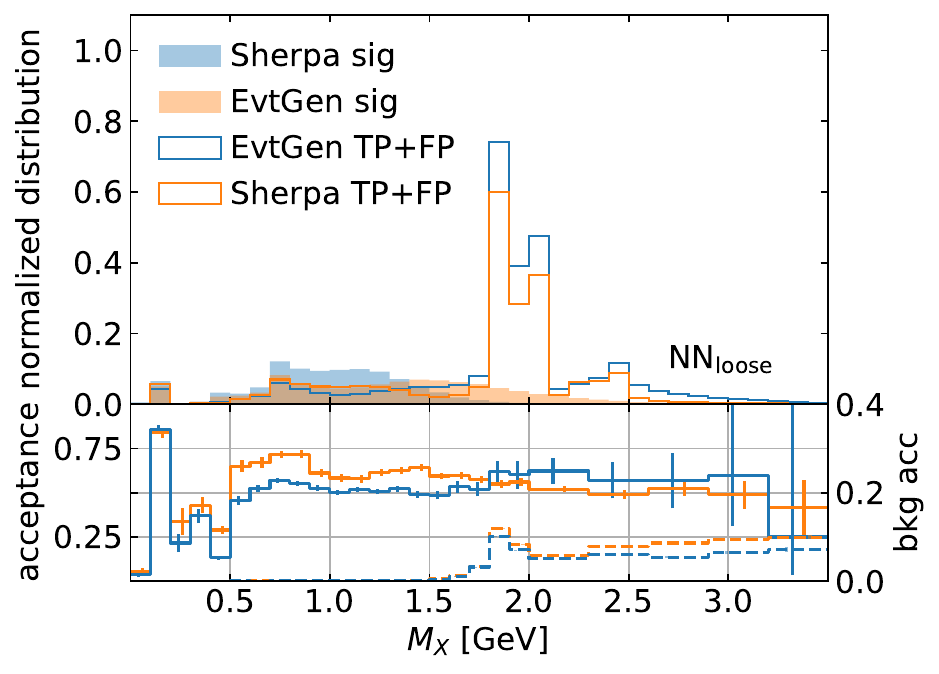}
    \;
    \includegraphics[height=5.2cm]{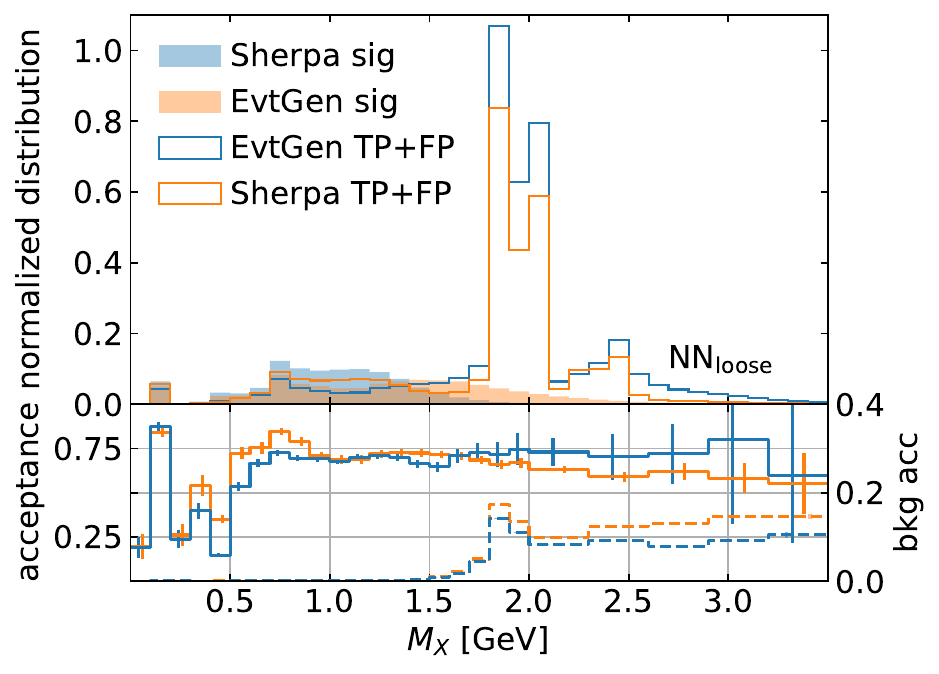}
  \caption{$M_X$ distributions and signal acceptance for \NNone (top) and \NNtwo (bottom) trained on \Evtgen (left) and \Sherpa (right) data. For \NNtwo the dashed lines in the lower panel show the background acceptance using the scale for the $y$-axis on the right.}The distributions in the upper panels of each plot are normalized  to the total number of signal events. A broader binning has been chosen to show the acceptance at $M_X>2\,\gev$, where event statistics are low. 
  \label{fig:sherpa_trained_vs_evtgen_trained}
\end{figure}

In Section~\ref{sec:inclusivity} we studied the performance of \NNone and \NNtwo when trained 
on \Evtgen data and then tested on both \Sherpa and \Evtgen data.  Here we give results 
when instead \Sherpa data is used to train the NNs.

We begin by showing in Figure~\ref{fig:sherpa_trained_vs_evtgen_trained} the signal acceptances of \NNone (upper row) and \NNtwo (bottom row), finely binned in the variable $M_X$. As in 
Figure~\ref{fig:high_extra_acceptance},  the plots also show the signal and total number of 
accepted events (TP+FP), normalized to the number detector-level signal events, in addition to 
the background acceptances for \NNtwo using the $y$-axis shown on the right of the lower panels.  
The plots in  the left-hand side of the figure are trained on \Evtgen data, while those on the right are trained on \Sherpa data.  

The figure shows that the signal acceptances for \NNone are fairly independent of the training and 
testing data up until about $M_X\sim 1.5$~GeV, even though finely-binned signal modelling from the two MCs is vastly different.  For $M_X>1.5$~GeV, on the other hand, the acceptances depend
crucially on the which MC is used in the training. The reason is that the \Sherpa signal drops quickly to zero beyond this point, and is already negligible at the $D$-meson resonance at $M_X=1.9$~GeV. 
Consequently, as seen in the top-right plot, a \Sherpa-trained \NNone tends to reject the higher-$M_X$ region of the \Evtgen signal, as it has not seen signal events in that region during the training.   

This artificial separation of signal and background in \Sherpa is an unphysical effect that can be remedied by a matching with OPE-based results, which give a model-independent description of 
fully inclusive rates in the higher-$M_X$ region.  We note further that  the signal acceptance 
of \NNtwo is fairly flat as a function $M_X$, whether trained on \Evtgen or
\Sherpa data, and in particular even the \Sherpa-trained version accepts \Evtgen signal events across the entire region. In this case, however, the unphysical behaviour of the signal modelling would inevitably show up in a poor fit quality in the second stage of the analysis.  For these reasons we have not considered \Sherpa-trained NNs in the body of the text. 

\begin{figure}[tbh]
\centering
	\includegraphics[height=5.2cm]{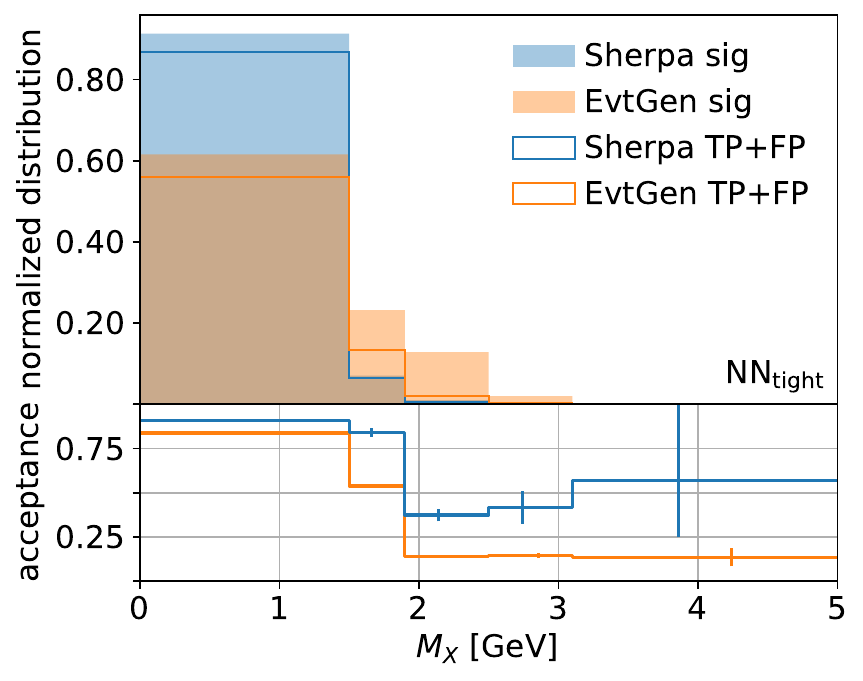} \quad 
\includegraphics[height=5.2cm]{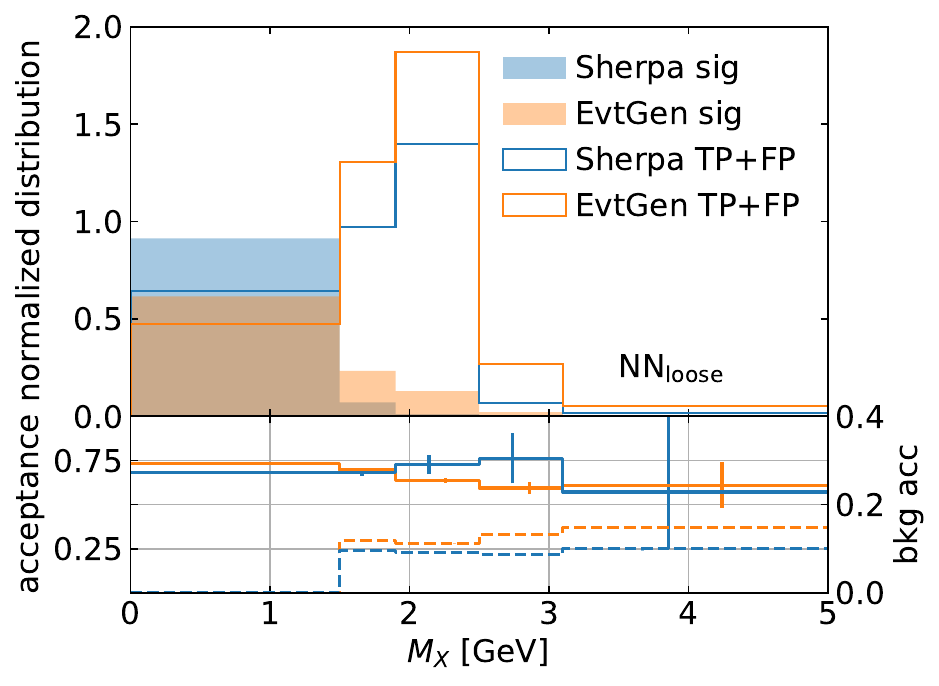}
\\[2mm]
\includegraphics[height=5.2cm]{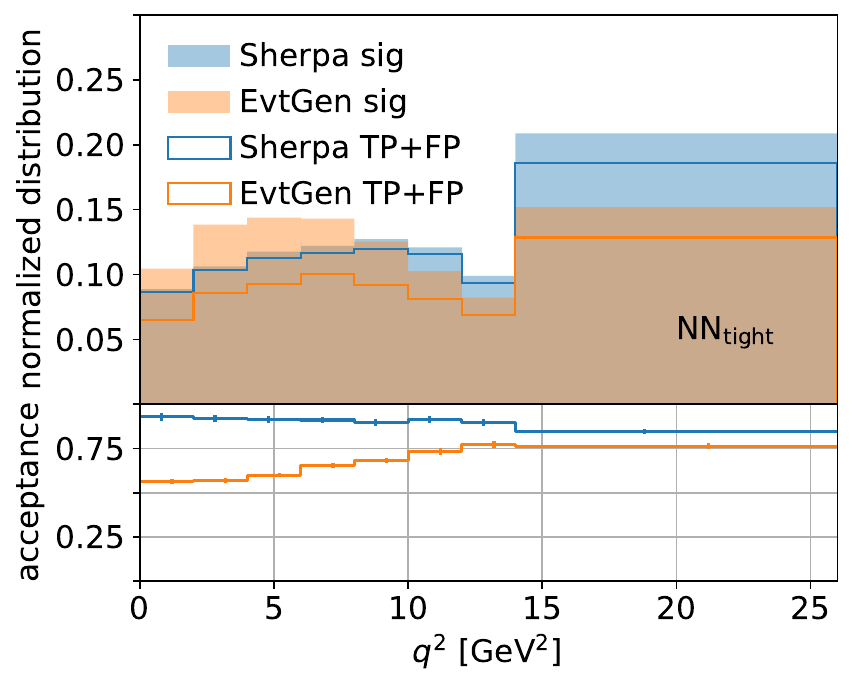}\quad
	\includegraphics[height=5.2cm]{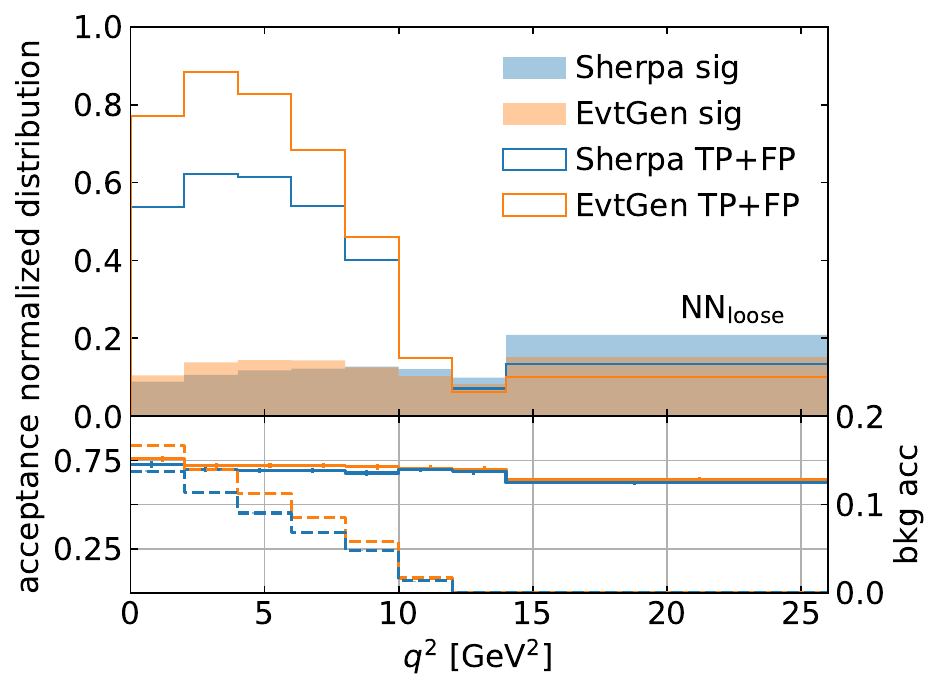}
  \\[2mm] 
   	\includegraphics[height=5.2cm]{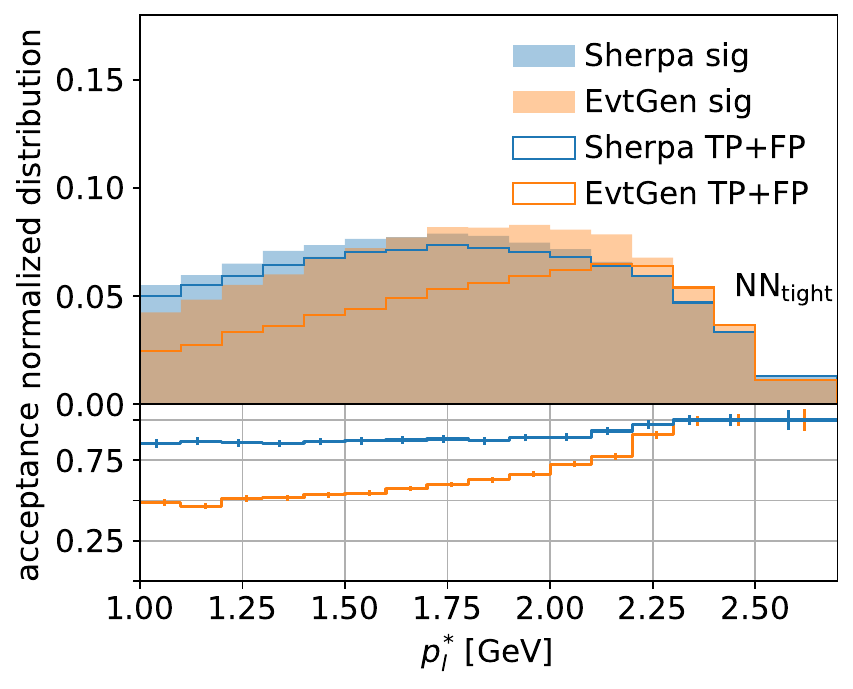}\quad
	\includegraphics[height=5.2cm]{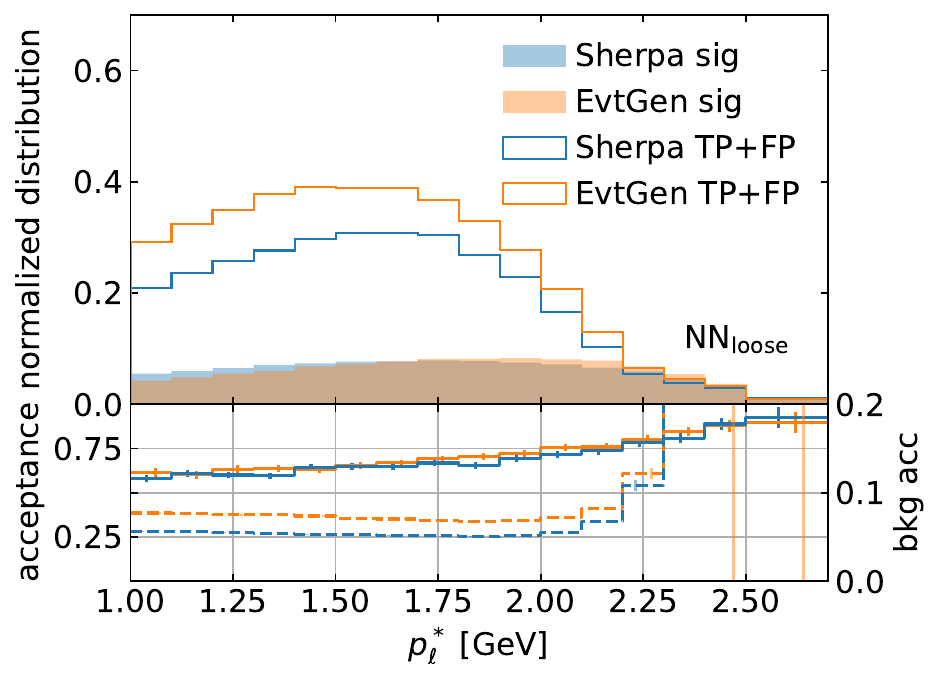}
    \\[2mm]
	\caption{As in Figure~\ref{fig:high_extra_acceptance}, but using \Sherpa instead of \Evtgen data for training the BNNs. }
	\label{fig:high_extra_acceptance_sherpa}
\end{figure}

\begin{figure}[thb]
  \centering
    \includegraphics[height=5.5cm]{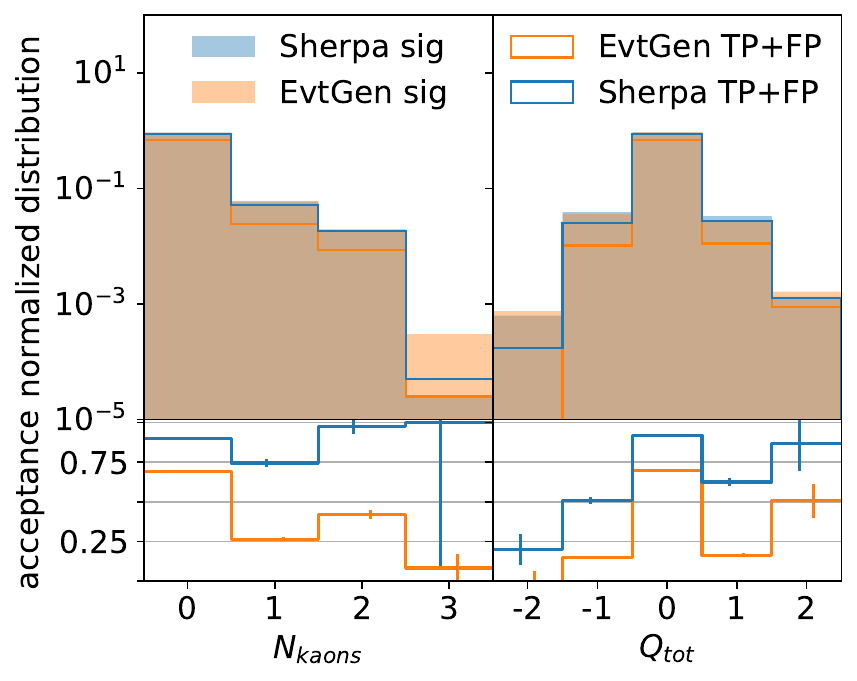}\;
    \includegraphics[height=5.5cm]{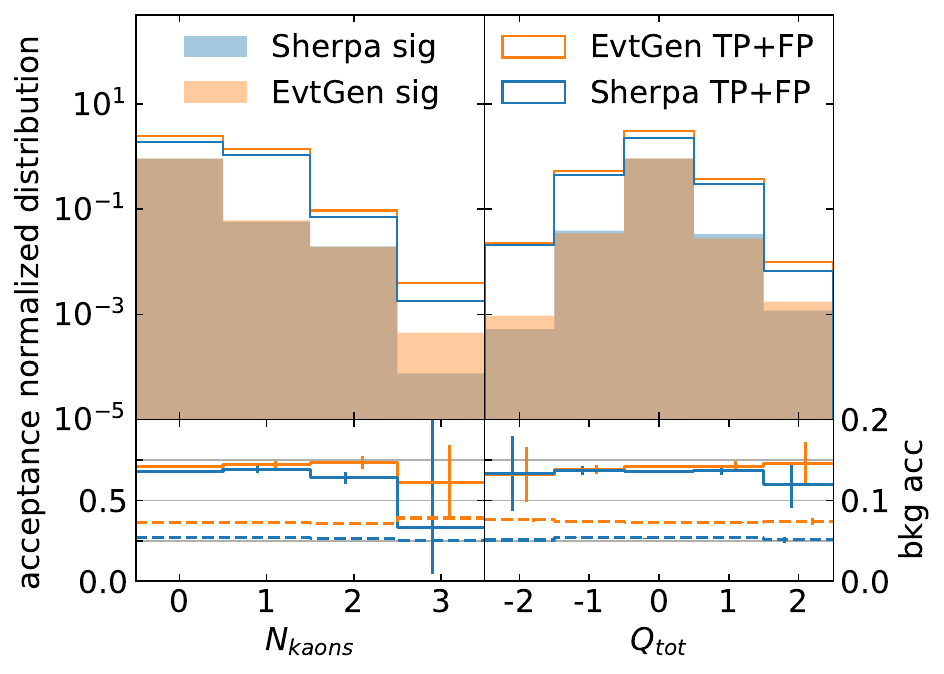}\\
  \caption{As in Figure~\ref{fig:sherpa_trained_vs_evtgen_trained_global}, but using \Sherpa instead of \Evtgen data for training the BNNs.
  }
  \label{fig:sherpa_trained_trained_global}
\end{figure}

Still, for completeness, we show in Figures~\ref{fig:high_extra_acceptance_sherpa} and~\ref{fig:sherpa_trained_trained_global} the \Sherpa-trained versions of Figures~\ref{fig:high_extra_acceptance} and~\ref{fig:sherpa_trained_vs_evtgen_trained_global}.    The most prominent feature is 
the expected reduction in the signal acceptance of \Evtgen data by \NNone in the regions of high-$M_X$ and low $q^2$ and $p_\ell^*$ in Figure~\ref{fig:high_extra_acceptance_sherpa} compared to the 
\Evtgen-trained version in Figure~\ref{fig:high_extra_acceptance}, as well as a higher acceptance of the \Sherpa signal overall,  regardless of the NN. 

\FloatBarrier
\newpage
\bibliographystyle{JHEP.bst}
\bibliography{literature}

\end{document}